\documentclass[aps,pra,reprint,groupedaddress, showpacs,amssymb,nofootinbib,]{revtex4-1}
\usepackage{amsmath,amssymb}
\usepackage{bm}
\usepackage{epsfig}
\numberwithin{equation}{section}
\renewcommand{\theequation}{\arabic{section}.\arabic{equation}}
\def\erf{\text{erf}}
\usepackage{pstricks-add,pst-plot}
\usepackage{enumerate}
\usepackage{braket}
\usepackage{relsize}
\usepackage{hyperref}
\hypersetup{
colorlinks = true,
linkcolor = blue,
citecolor = blue,
urlcolor = blue,
}
\everymath{\displaystyle}
\begin{document}

\title{Decoherent histories for a particle coupled to a von Neumann apparatus}

\author{Francesc S. Roig}
\email{roig@physics.ucsb.edu}

\affiliation{Physics Department University of California Santa Barbara, CA 93106, USA}

\date{\today}


\begin{abstract}
{\label{abstract}}
 Using the Gell-Mann and Hartle formalism of generalized quantum mechanics of closed systems, we study coarse-grained decoherent histories. The system under consideration is one-dimensional and consists of a particle coupled to a von Neumann apparatus that measures its position.  The particle moves in a quadratic potential; in particular we consider a driven harmonic oscillator. The real line is divided into intervals of the same length, and coarse-grained histories are defined by the arithmetic average of the initial and final position of the particle to be within one of these intervals. The position of the pointer correlates with this arithmetic average. In addition a constant term is added to this average, which is a result of the presence of a driving force on the oscillator.   We investigate decoherence for such histories via the decoherence functional for the particle-apparatus. An exact expression for this functional has been derived.  If the particle or the pointer of the apparatus is in an exact position state initially, then  exact decoherence ensues, and the relative probability for such histories is obtained. If the initial state of the particle, which is centered at some arbitrary point in the $x$-axis, is  narrow compared to the width of the intervals on the~$x$-axis,  then we obtain approximate decoherence. The same result follows if the initial state of the pointer is narrow. In these two situations, we evaluate expressions for the decoherence functional qualitatively and quantitatively. In the latter case we obtain the first two terms in an expansion in powers of the width of the initial state of the particle or the pointer respectively.  Approximate relative probabilities can be assigned in this case, and we have obtained an expansion up to second order in powers of the width of the initial state for an expression for an upper bound of these probabilities.
\end{abstract}

\maketitle


\section{Introduction}\label{Introduction}
\label{intro}
In this paper we consider the system of a particle interacting with a von Neumann measuring apparatus, which was studied in detail in references~\cite{FR,FR-1}.  Standard quantum mechanics was applied to this system to study quantum measurements of finite duration of the position of the particle.   

In the current work we set out to investigate questions about decoherence in the sense of the Gell-Mann and Hartle formalism of generalized quantum mechanics of closed systems~\cite{H1,H2} when applied to the system of a particle coupled to a von Neumann apparatus that measures the position of the former. 

The decoherence functional in Refs.~\cite{H1,H2} is defined by a functional ~$D(c_{\alpha'},c_\alpha)$, where~$c_\alpha$ is a class of suitably defined coarse-grained histories.  This decoherence functional provides the following condition for exact decoherence:
\begin{equation}\label{medium decoherence}
D(c_{\alpha},c_{\alpha'})=0,\quad\alpha'\not =\alpha.
\end{equation}
This condition is also known as medium decoherence in Ref.~\cite{G-MH}. 
When  a set of histories decoheres  probabilities can be assigned to such histories.
Presently the decoherence functional will be evaluated to the extent needed to obtain meaningful results and conclusions  when applied to the system particle-apparatus, where the position of the particle is measured by a von Neumann apparatus. 
The question that we set out to answer is this:  For what types of the initial state of the system particle-apparatus can probabilities be assigned to coarse-grained histories that start at time~$t=0$ and end at~$t=T$?  Since the particle will be coupled to the apparatus, a finite time measurement of the position of the particle will have taken place.  That is, the pointer of the apparatus will have correlated with an indicator function which depends on the initial and final positions of the particle corresponding to the interval~$[0,T]$. Another way to express this is by saying that a "measurement situation" in the sense of Refs.~\cite{H2,G-MH} has taken place. Then, as pointed out in~\cite{H2} for the general case of measured operators, the measured position of the particle is correlated with decoherent histories for the system particle-apparatus, so that probabilities can be assigned to this measurement. 
 
 The criterion for coarse-grained histories will be established as follows: The real line will be divided up into contiguous intervals of the same length, and a coarse-grained history will consist of the  set of all spacetime histories in the time interval~$[0,T]$ whose arithmetic average of the initial and final position of the particle lies in a given interval.  Finally, for review works on decoherence in quantum mechanics see references~\cite{{Dowker-Kent},{Schlosshauer},{Dowker-Halliwell},{Halliwell}}.

The Hamiltonian for this system is
\begin{equation} \label{H}
H=\frac{p^2}{2m}+V(x,t)+\frac{P^2}{2M}+H_i.
\end{equation}
The mass of the particle is~$m$ and the mass of the apparatus or pointer is~$M$. The total potential acting on the particle is~$V(x,t)$.

The interaction between the particle and the apparatus is described by the von Neumann Hamiltonian~\cite{vonN} for the measurement of the position of a particle
\begin{equation} \label{Hi}
H_i = \frac{1}{T} f(t) x\hat P,
\end{equation} 
where
\begin{equation}\label{pointer_momentum}
\hat P=-{\rm i}\hbar\frac{\partial}{\partial X}
\end{equation}
 is the momentum of the pointer,with~$X$ the pointer position, or indicator variable. The duration of this interaction is~$T$. The function~$f(t)$ is a dimensionless coupling function of time with compact support~$[0,T]$. For the purposes of the current work, this function is assumed to be symmetric about the midpoint of the time interval~$[0,T]$. Just before the measurement, the system is described by the pure state $\psi_0(x, X) = \varphi_0(x) \Phi_0(X)$, where $\varphi_0(x)$ is the state of the particle and~$\Phi_0(X)$ can be a narrow wavepacket describing the apparatus.
 
In reference~\cite{FR} it was shown that the propagator for a particle-pointer system, with~$\hbar=1$,  is written as a sum over all paths between~$0$ and~$T$ as
\begin{equation} \label{2D-pathint}
\left\langle {x,X} \right|e^{ - {\rm i}HT} \left| {x',X'} \right\rangle  = \iint\delta x(t)\delta X(t)e^{{\rm i}S[x(t),X(t)]},
\end{equation}
where the action is
\begin{align}\label{S}
S\left[ {x(t),X(t)} \right] = &\mathlarger{{\mathlarger{\int}}}_{\!\!\!\!\!0}^T\Bigg[ \frac{m}{2}\dot x^2  - V(x,t)\nonumber\\
&  
 +\frac{M}{2}\left( {\dot X - \frac{f(t)}{T}x} \right)^2\Bigg] {\rm d}t.
\end{align}

The path integral over all pointer paths~$X(t)$~in \eqref{2D-pathint} can be readily evaluated:
change the pointer variable according to
\begin{equation}
Y=X-\frac{1}{T}\int_0^t f(t')x(t'){\rm d}t'.
\end{equation}
With this change the double path integral~in \eqref{2D-pathint} can be rewritten
\begin{align} \label{1D-pathint}
&\mathlarger{\iint} \delta x(t)\delta X(t)e^{{\rm i}S[x(t),X(t)]}=\nonumber\\
&\mathlarger{\mathlarger{\int}} \delta x(t)~{\rm exp}\left\{{\rm i}\int_0^T{\rm d}t \left[\frac{m}{2}\dot x^2
-V(x,t)\right] \right\}\nonumber\\
&\times
\mathlarger{\mathlarger{\int}} \delta Y(t)~{\rm exp}\left({\rm i}\int_0^T {\rm d}t\frac{M}{2}\dot Y^2\right).
\end{align}
The path integration over~$Y(t)$ yields the result for a free particle of mass~$M$
\begin{align}
\mathlarger{\mathlarger{\int}}  \delta &Y(t)~{\rm exp}\left({\rm i}\int_0^T {\rm d}t\frac{M}{2}\dot Y^2\right)=\nonumber\\
&
\sqrt{\frac{M}{2\pi iT}}~{\rm exp}\left[{\rm i}\frac{M}{2T}\left(Y'-Y\right)^2\right].
\end{align}
That is, the propagator~\eqref{2D-pathint} can be expressed in terms of a single path integral
over all particle paths
\begin{align}\label{X-pathint}
&\Braket{x,X|{\rm e}^{-{\rm i}HT}|x',X'} =\nonumber\\
&
\mathlarger{\mathlarger{\int}} \delta x(t)~{\rm exp}\left[{\rm i}\int_0^T {\rm d}t\left(\frac{m}{2}\dot x^2-V(x,t)\right)\right]\nonumber\\
&\times 
\sqrt{\frac{M}{2\pi {\rm i} T}}
~{\rm exp}\left\{{\rm i} \frac{M}{2T}\Big[X-X'-\bar x[x(t),T]\Big]\right\}
\end{align}
where
\begin{equation}\label{bar x}
\bar x[x(t)]=\frac{1}{T}\int_0^T f(t) x(t){\rm d}t
\end{equation}
is the  average of~$x(t)$ on a particle path weighted by the coupling function~$f(t)$.

If initially the particle is in the pure state~$\psi_0(x,X)=\varphi_0(x)\Phi_0(X)$, then
the wavefunction for the system at time~$T$ is given by the entangled state
\begin{align}\label{FinalStateT}
\psi&(x,X,T)=\int_{-\infty}^\infty{\rm d}x'\varphi_0(x')\nonumber\\
&\times
\mathlarger{\mathlarger{\int}}_{(x',0)}^{(x,T)}\delta x(t)
{\rm exp}\left\{{\rm i}\int_0^T\left[\frac{m}{2}\dot x^2-V(x,t)\right]{\rm d}t\right\}\nonumber\\
&\times
\Phi\left(X-\bar x[x(t)],T\right)
\end{align}

where
\begin{align}\label{Phi0}
&\Phi\left(X-{\bar x}[x(t)],T\right)=\sqrt{\frac{M}{2\pi {\rm i} T}}\nonumber\\
&\times
\mathlarger{\mathlarger{\int}}_{-\infty}^\infty {\rm d}X'\Phi_0(X')
{\rm exp}\left\{{\rm i} \frac{M}{2T}\Big(X-X'-\bar x[x(t)]\Big)^2\right\}.
\end{align}
Expression~\eqref{Phi0} shows that on a given Feynman path for the particle, the wavepacket describing the pointer at~$t=0$ and centered at~$X=0$ has spread at ~$t=T$ and shifted its center to~$\bar x [x(t)]$.  The pointer coordinate correlates with~$\bar x$ on an individual Feynman path.
 
For a particle in an eigenstate of position at~$x_0$ at~$t=0$ the wavefunction of the system at the end of the measurement is
\begin{align}\label{SharpParticleState}
\psi(x,X,T)=&
\mathlarger{\mathlarger{\int}}_{(x_0,0)}^{(x,T)}\delta x(t)
{\rm exp}\Bigg\{{\rm i}\int_0^T\Big[\frac{m}{2}\dot x^2\nonumber\\
&
-V(x,t)\Big]{\rm d}t\Bigg\}
\Phi\left(X-\bar x[x(t)],T\right),
\end{align}
 the path integral sums over all particle paths that start at~$x_0$ at~$t=0$ and end at any point on the real line at~$t=T$.

For the case of quadratic particle potentials
\[
V(x,t)=V_0(t)+V_1(t)x+V_2x^2
\]
with~$V_2={\rm Constant}>0$, the propagator~\eqref{2D-pathint} factorizes in the form:
\begin{align}\label{PropQPot}
&\Braket{x,X|e^{{- \rm i}HT}| x',X'}\!=\!\Braket{x|{\rm e}^{-{\rm i}H_0T}|x'}{\rm e}^{{\rm i}\phi(x,x')}
\nonumber\\
\times &
\sqrt{\frac{M_\mathit {eff}}{2\pi{\rm i}T}}~{\rm exp}\bigg\{\frac{{\rm i}M_\mathit {eff}}{2T}\big[X-X'-s(x,x')\big]^2\bigg\}
\end{align}
where
\[
\Braket{x|{\rm e}^{-{\rm i}H_0T}|x'}=\mathlarger{\int}\delta x(t){\rm e}^{{\rm i}S_0[x(t)]},
\]
and
\begin{equation}\label{S0}
S_0[x(t)]=\mathlarger{\int}_0^T{\rm d}t\bigg[\frac{m}{2}\dot x^2-V_2x^2-V_0(t)\bigg].
\end{equation}

For a driven oscillator acted on by a driving force~$f_D(t)$ symmetric about the midpoint of the interval~$[0,T],$ the expressions for~$s(x,x')$,~$M_{\mathit{eff}}$ and~$\phi(x,x')$ in~\eqref{PropQPot}, are given by equations (29), (31) and (34) in Ref.~\cite{FR-1} respectively. In particular in Ref.~\cite{FR-1} the shift function takes the form 
\begin{equation}\label{shiftfunction}
s(x,x')=g(\omega,T)\left(\frac{x+x'}{2}\right)+d(\omega,T).
\end{equation}

When the initial state of the particle is a sharp state of position, then the wavefunction is of the form~$\psi_0(x,X,T)=\delta(x-x_0)\Phi_0(X)$, where~$\Phi_0(X)$ is a wavepacket centered a~$X=0$ describing the initial state of the pointer. The wavefunction of the system at time~$T$ is
\begin{equation}
\psi(x,X,T)=\Phi\big(X-s(x_0,x),T\big)\mathlarger{\int}\delta x(t){\rm e}^{{\rm i}S_0[x(t)]}
\end{equation}
where
\begin{align}\label{Phi(X,T)}
\Phi\big(&X-s(x_0,x),T\big)=\sqrt{\frac{M_\mathit {eff}}{2\pi{\rm i}T}}\mathlarger{\mathlarger{\int}}_{-\infty}^\infty{\rm d}X'\nonumber\\
&\times
{\rm exp}\bigg\{\frac{{\rm i}M_{\mathit{eff}}}{2\pi{\rm i}T}\big[X-X'-s(x_0,x)\big]^2\bigg\}\Phi_0(X').
\end{align}
Thus the indication of the apparatus correlates with the arithmetic average of the initial and final positions.

In Section \ref{formalism} we will introduce coarse-grained histories for the system particle-apparatus and we will apply the quantum mechanics of closed systems~\cite{H1,H2} to these histories.  We will find that exact probabilities can be assigned whenever the particle starts at an eigenstate of position, and likewise for the pointer.  When the initial states of the particle or the pointer are narrow there will be approximate decoherence.  
In this section we show the basic development as well as presenting an exact expression for the decoherence functional. 

 The basic formalism developed in Sect.~\ref{formalism} is applied in Section~\ref{exact decoherence} to show that there is exact decoherence for systems starting in an eigenstate of position either for the particle or for the pointer. Subsequently exact probabilities will be  evaluated for the coarse-grained histories.  In Section~\ref{Section IV} semi-qualitative calculations for the decoherence functional and for an upper bound for approximate probabilities are presented.  
Sections~\ref{Section V} and~\ref{Section VI} develop detailed calculations of the decoherence functional and upper bound for the probabilities for the cases of a narrow Gaussian initial particle state and a narrow initial Gaussian state for the pointer respectively.  Section~\ref{conclusion} is the summary and conclusion.  


\section{Formalism}
\label{formalism}
In this section we develop the basic formalism on how to apply the quantum mechanics of closed systems to a particle coupled to a von Neumann apparatus. To this end the set of all particle paths~$C$ starting on any point on the real line ($x$-axis) at~$t=0$ and ending on any point on the real line ($x'$-axis) at~$t=T$ can be grouped into an exhaustive set of mutually exclusive classes.  Consequently, as in reference~\cite{B-H}, the real line of all values for the average position~$\bar x$ on a given Feynman path for the particle is divided into equal length intervals:
\begin{align}
\Delta_{\alpha}&=\big(\,\bar x_{\alpha}-\delta/2\,,\,\bar x_{\alpha}+\delta/2\,\big],~~{\rm with}~\delta>0\nonumber\\
\mathbb{R}&=\bigcup_{\alpha\in \mathbb{Z}}\Delta_{\alpha}\,,~{\rm with}~~ \bar x_{\alpha+1}-\bar x_\alpha=\delta.
\end{align}
A class of particle paths~$c_{\alpha}$ is defined by the set of all paths~$x(t)$ on the time interval~$[0,T]$  for which on any given path we get the average~$\bar x$ such that~$\bar x\in \Delta_{\alpha}$.
Therefore  the set~$C$ of all paths in the time interval~$[0,T]$  has been divided into an exhaustive set of mutually exclusive classes~$c_{\alpha}$:
\begin{align}
C&=\bigcup_{\alpha\in \mathbb{Z}}c_{\alpha}\nonumber\\
\O&=c_{\alpha}\bigcap c_{\alpha'},~~{\rm with}~\alpha\not =\alpha'.
\end{align}

There are two possibilities on how the average of~$x$ is to be taken:
\begin{enumerate}[A)]
\item
The average of~$x$ is given by~\eqref{bar x}, the continuous weighted average of the position of the particle  over a Feynman path.
\item
The average of~$x$ is given by the arithmetic average of the initial and final position of the particle~$\bar x=(x+x')/2$.
\end{enumerate}

 Alternative (A) is suitable for any potential~$V(x,t)$ acting on the particle.  In this case for each particle path in a class~$c_{\alpha}$, the pointer correlates with~$\bar x[x(t)]\in\Delta_\alpha$.
That is the pointer indicator is somewhere within the interval~$\Delta_\alpha$. This case requires a separate study which would fall outside the scope of the present work. However, the interesting properties of decoherence do not seem to occur in this case for the particle-apparatus system, starting with simple situations, namely: a free particle and an initial state of the system that is an eigenstate of position either for the particle or for the pointer.  Not even the weak decoherence condition
\begin{equation*}
{\rm Re}[D(c_{\alpha},c_{\alpha'})]=0\quad{\rm for}\quad\alpha\neq\alpha'
\end{equation*}
is satisfied in this case.

Case (B) can be used when the potential of the particle is quadratic, and this will be the situation considered for the rest of this work. In this case for each particle path in~$c_{\alpha}$, the pointer correlates with the shift function~$s(x,x')$. For driven oscillators, when the driving force~$f_{D}(t)$ is symmetric about the midpoint of~$[\,0\,\,T\,]$, the shift function takes the form~\eqref{shiftfunction}.
The fine-grained histories for this system are given by any particle and pointer paths~$\big(x(t),X(t)\big)$ defined on the time interval~$[0,T]$. The coarse-grained-classes  consist of all the fine-grained histories belonging to the set
\begin{equation}\label{histories}
c^{system}_{\alpha}=\Set{\big(x(t), X(t)\big) | \bar x\equiv\frac{x+x'}{2}\in \Delta_{\alpha}}.
\end{equation}
Thus all the fine-grained histories  for the particle with the same value~$\bar x\in \Delta_\alpha$ are contained in~$c_{\alpha}$

The essential feature of the quantum mechanics of closed systems~\cite{{H1},{H2}}, is that for each class of paths~$c_{\alpha}^{\,system}$, a class operator~$\widehat C_{\alpha}^{\,system}$ can be defined whose matrix elements in the position representation takes the form of a path integral unrestricted over all pointer paths and restricted to all possible particle paths in~$c_{\alpha}$
\begin{align}\label{class-operator}
&\Braket{x,X|\widehat C_{\alpha}^{\,system}|x',X'}=\nonumber\\
&
\int\limits_{(X',0)}^{(X,T)}\delta X(t)
\int_{c_{\alpha}}\delta x(t) \,{\rm e}^{{\rm i} S[X(t),x(t)]},
\end{align}
where the action~$S$ is given by~\eqref{S}. For a driven oscillator the potential is 
\begin{equation}\label{V driven SHO}
V(x,t)=\frac{m}{2}\omega^2x^2-f_D(t)x.
\end{equation}

When the system starts at~$t=0$ in the state described by the state vector~$\Ket{\Psi_0}$, the decoherence functional is defined by
\begin{equation}\label{decoherence-functional}
D(c_{\alpha},c_{\alpha'})=\Braket{\Psi_0|(\widehat C^{\,system}_{\alpha'})^{\dagger}\,\,\widehat C^{\,system}_\alpha|\Psi_0}.
\end{equation}
The quantum mechanics of closed systems does not predict probabilities for every possible set of coarse-grained histories but only for those that satisfy the decoherence condition
\begin{equation}\label{decoherence}
D(c_{\alpha},c_{\alpha'})\thickapprox 0,\quad\alpha'\not =\alpha.
\end{equation}

In such a case, probabilities for each of the coarse grained classes~$c^{system}_{\alpha}$ can be assigned according to  
\begin{equation}
p_\alpha=\parallel\widehat C^{\,system}_{\alpha}\Ket{\Psi_0}\parallel^2.
\end{equation}

The calculation of the matrix elements of the class operators~$\widehat C_\alpha$ is straightforward for quadratic particle potentials.  We begin with the definition of the matrix elements of~$\widehat C_\alpha$ in~\eqref{class-operator}.  The path integral can be rewritten introducing the function
\begin{equation}\label{top-hat function}
\mathlarger{e}_{\mathsmaller{\Delta}_{\alpha}}(\bar x)=
\begin{cases}
\,\,1,\quad \bar x\in\Delta_\alpha\\
\,\,0,\quad \bar x\notin\Delta_\alpha\,.
\end{cases}
\end{equation}
Thus we may rewrite the functional integral over the class of paths~$c_\alpha$ in Eq.\eqref{class-operator} as
\begin{align}\label{particle-path-int}
\int_{c_{\alpha}}&\delta x(t)\,{\rm e}^{{\rm i} S[X(t),x(t)]}=\nonumber\\
&\mathlarger{e}_{\mathsmaller{\Delta}_{\alpha}}\left(\frac{x+x'}{2}\right)
\mathlarger{\int}\limits_{(x',0)}^{(x,T)}\delta x(t){\rm e}^{{\rm i}S[X(t),x(t)]}\,.
\end{align}
Inserting Eq.~(\ref{particle-path-int}) into Eq.~(\ref{class-operator}) we obtain a simple relation for the matrix elements of the class operators and the propagator~\eqref{2D-pathint} of the system
\begin{align}\label{class-op-matrix-elements1}
&\Braket{x,X|\widehat C_{\alpha}^{\,system}|x',X'}=\nonumber\\
& \mathlarger{e}_{\mathsmaller{\Delta}_{\alpha}}\left(\frac{x+x'}{2}\right)
\left\langle {x,X} \right|{\rm e}^{ - {\rm i}HT} \left| {x',X'} \right\rangle. 
\end{align}

The next object of interest is the branch wavefunction defined as
\begin{equation}\label{branch-wavefunction1}
\Psi_\alpha(x,X,T)=\Braket{x,X|\widehat C^{\,system}_\alpha|\Psi_0}.
\end{equation}
The branch wavefunction may be rewritten in terms of the matrix elements~\eqref{class-operator} for the class operators
\begin{align}\label{branch-wavefunction2}
\Psi_\alpha(x,X,T)=&\mathlarger{\int}_{-\infty}^\infty \!\!\!{\rm d}X'\mathlarger{\int}_{-\infty }^\infty\!\!\!\Braket{x,X|\widehat C_\alpha^{\,system}|x',X'}\nonumber\\
&\times
\Psi_0(x',X'){\rm d}x'\,.
\end{align}
Inserting~\eqref{class-op-matrix-elements1} into the expression~\eqref{branch-wavefunction2} for the branch wavefunction and observing that the  limits for the integration over~$x'$ are determined by the function~\eqref{top-hat function}, or equivalently 
\begin{equation}\label{limits}
\bar x_\alpha-\frac{\delta}{2}\le\frac{x+x'}{2}< \bar x_\alpha+\frac{\delta}{2}
\end{equation}
we obtain
\begin{align}\label{branch-wavefunction3}
\Psi_\alpha (x,X,T)=&\mathlarger{\int}_{-\infty}^\infty{\rm d}X'
\mathlarger{\int}_{2\bar x_{\alpha}-x-\delta}^{2\bar x_\alpha-x+\delta}{\rm d}x'\nonumber\\
&\times
K_{TOT}(x,X,T;x',X',0)\nonumber\\
&\times
\Psi_0(x',X')
\end{align}
with~$K_{TOT}(x,X,T;x',X',0)$ the propagator of the system. That is,
\begin{equation}\label{TOT-Prop}
K_{TOT}(x,X,T;x',X',0)=
\Braket{x,X|{\rm e}^{-{\rm i}HT}|x',X'}
\end{equation}
and
\begin{equation}
\Psi_\alpha(x,X,T)=0,\quad {\rm if}\quad \frac{x+x'}{2}\notin\Delta_\alpha\,.
\end{equation}
The decoherent functional~\eqref{decoherence-functional} can be rewritten in terms of the branch-wavefunction
\begin{equation}\label{decoherence-functional1}
D(c_\alpha,c_{\alpha'})=\!\!\int\limits_{\!
\!\!\!\!\!\!\!\!-\infty}^{\,\,\infty}\int\limits_{\!\!\!\!\!-\infty}^{\,\,\,\,\infty}\!\!{\rm d}X{\rm d}x\,\Psi^{*}_{\alpha'}(x,X,T)\Psi_{\alpha}(x,X,T),
\end{equation}
and exact decoherence can be expressed as the orthogonality of the branch wavefunctions
\begin{equation*}
\Braket{\Psi_{\alpha'}|\Psi_\alpha}=0.
\end{equation*}
In this case the probability for the alternative~$\alpha$ is
\begin{equation*}
p_\alpha=\Braket{\Psi_{\alpha}|\Psi_\alpha}.
\end{equation*}
Substituting~\eqref{branch-wavefunction3} into the expression~\eqref{decoherence-functional1} we obtain the decoherence functional expressed in terms of the propagator of the system and the initial state wavefunction of the system
\begin{align}\label{decoherence-functional2}
D(c_{\alpha},c_{\alpha'})&=\iint\limits_{\!\!\!\!\!-\infty}^{\,\,\,\,\,\,\,\,\,\infty}{\rm d}X{\rm d}x\iint\limits_{\!\!\!\!\!-\infty}^{\,\,\,\,\,\,\,\,\,\infty}{\rm d}X''{\rm d}X'\nonumber\\
&\times
\mathlarger{\int}_{2\bar x_{\alpha'}-x-\delta}^{2\bar x_{\alpha'}-x+\delta}{\rm d}x''
\mathlarger{\int}_{2\bar x_{\alpha}-x-\delta}^{2\bar x_{\alpha}-x+\delta}{\rm d}x'\nonumber\\
&\times
K^{*}_{TOT}(x,X,T;x'',X'',0)\nonumber\\
&\times
K_{TOT}(x,X,T;x',X',0)\nonumber\\
&\times
\Psi_0^{*}(x'',X'')\Psi_0(x',X').
\end{align}

Next we will obtain and alternative expression for the decoherent functional and we will assume that the initial state is a pure state of the form~$\Psi_0(x,X)=\varphi_0(x)\Phi_0(X)$.
In reference~\cite{FR-1} the following expression was obtained for the propagator of a driven harmonic oscillator interacting with a von Neumann apparatus:
\begin{widetext}
\begin{align} \label{PropTOT1}
K_{TOT}(x,X,T;x',X',0) =&K_{0}(x,T;x',0)\exp\left[{\rm i}\phi\left(\frac{x+x'}{2},\omega,T\right)\right] \nonumber\\
&\times
\left(\frac{M_\mathit{eff} }{2\pi {\rm i}T}\right)^{1/2}\exp \left\{ {{\rm i}\frac{M_\mathit{eff}}{2T}\left [ {X - X' - g(\omega ,T)\left( {\frac{x + x'}{2}} \right) - d(\omega ,T)} \right ]^2} \right\},
\end{align}
\end{widetext}
where~$M_{\mathit{eff}}$ and the phase~$\phi$ in~\eqref{PropTOT1} are given in~\cite{FR-1} by Eq.(29) and Eq.(33) respectively.  Likewise in~\cite{FR-1} the coupling constant~$g(\omega,T)$ and the displacement~$d(\omega,T)$ are given by Eq.(31) and Eq.(32) respectively, and $K_0(x,T,x',0)$ is the propagator for the free harmonic oscillator.
In the appropriate limits Eq.\eqref{PropTOT1} reduces to the propagators for a free particle, a free harmonic oscillator~\cite{FR}, and the propagator for a particle acted on by a time-dependent force.

We begin by defining the unitary operator for a free pointer of mass~$M_{\mathit{eff}}$ and with~$\hat P$ given by~\eqref{pointer_momentum}
\begin{equation}
\hat U_{\mathit{eff}}(0,T)=\exp\left(-{\rm i}\frac{\hat P^2}{2M_{\mathit {eff}}}T\right)\,.
\end{equation}
Then in the expression~\eqref{PropTOT1} for the propagator of the system we can rewrite 
\begin{align}
&\Braket{X|\hat U_{\mathit{eff}}(T,0)|X'+s(x,x')}=\nonumber\\
&
\left(\frac{M_\mathit{eff} }{2\pi {\rm i}T}\right)^{1/2}\exp\left\{ {\rm i}\frac{M_\mathit{eff}}{2T}\left[X-X'-s(x,x')\right]\right\},
\end{align}
where~$s(x,x')$ is the shift function~\eqref{shiftfunction}.

In Eq.~\eqref{decoherence-functional2} the integrations over the pointer variables for the decoherence functional can be rewritten as shown below.
\begin{align}\label{int-pointer-variables}
&\mathlarger{\iiint}\limits_{\!\!\!\!-\infty}^{\,,\,\,\,\,\,\,\infty}{\rm d}X''{\rm dX'}{\rm dX}\,K^{*}_{TOT}(x,X,T;x'',X'',0)\nonumber\\
&\times
K_{TOT}(x,X,T;x',X',0)
\nonumber\\
&\times
\Psi_0^{*}(x'',X'')\Psi_0(x',X')\nonumber\\
&=\mathlarger{\iint}\limits_{\!\!\!\!-\infty}^{\,\,\,\,\,\,\,\,\,\,\infty}{\rm d}X''{\rm d}X'\Psi_0^*(x'',X'')\Psi_0(x',X')
\nonumber\\
&\times
\mathlarger{\int}_{-\infty}^\infty dX
\Braket{X''+s(x,x'')|\hat U^{\dagger}_{\mathit{eff}}|X}\nonumber\\
&\times
\Braket{X|\hat U_{\mathit{eff}}|X'+s(x,x')}.
\end{align}
Next the~$X$-integration can easily be carried out yielding a Dirac delta function
\begin{align}\label{X-int}
\mathlarger{\int}_{\infty}^\infty {\rm d}X&
\Braket{X''+s(x,x'')|\hat U^{\dagger}_{\mathit{eff}}|X}\nonumber\\
&\times
\Braket{X|\hat U_{\mathit{eff}}|X'+s(x,x')}=\nonumber\\
&
\delta\left(X''-X'+g(\omega,T)\left(\frac{x''-x'}{2}\right)\right).
\end{align}
Inserting~\eqref{X-int} into \eqref{int-pointer-variables} we obtain
\begin{align}\label{pointer-ints}
\mathlarger{\iiint}\limits_{\!\!\!\!-\infty}^{\,,\,\,\,\,\,\,\infty}{\rm d}X''{\rm d}X'&{\rm d}XK^{*}_{TOT}(x,X,T;x'',X'',0)\nonumber\\
&\times
K_{TOT}(x,X,T;x',X',0)\nonumber\\
&\times
\Psi_0^{*}(x'',X'')
\Psi_0(x',X')\nonumber\\
&
=\varphi_0(x'')\varphi_0(x')
\Delta_{\mathit{overlap}}\,,
\end{align}
where the overlap integral involving the pointer wavefunction at~$t=0$ is
\begin{align}\label{pointer-overlap}
\Delta_{\mathit{overlap}}=\mathlarger{\mathlarger{\int}}_{-\infty}^\infty&{\rm d}X'\Phi_0^{*}\left(X'-g(\omega,T)\frac{x''-x'}{2}\right)\nonumber\\
&\times
\Phi_0(X')\,.
\end{align}
If the initial state of the system shows entanglement between the particle and the pointer, then in~\eqref{pointer-ints} we replace~$\varphi^*(x'')\varphi(x')\Delta_{\mathit{overlap}}$ by~\eqref{pointer-overlap}, 
with~$\Psi_0(x,X)$ instead of~$\Phi_0(X)$.

Finally the decoherence functional can be rewritten in the alternative exact form after we insert~\eqref{PropTOT1},~\eqref{pointer-ints} and~\eqref{pointer-overlap} into~\eqref{decoherence-functional2}:
\begin{widetext}
\begin{align} \label{decoherence-functional3}
D(c_{\alpha},c_{\alpha'})=&\mathlarger{\mathlarger{\iiint}}\limits_{\!\!\!\!-\infty}^{\,,\,\,\,\,\,\,\infty}{\rm d}x\,{\rm d}x'\,{\rm d}x''\,\mathlarger{e}_{\mathsmaller{\Delta}_{\alpha'}}\left(\frac{x+x''}{2}\right)\mathlarger{e}_{\mathsmaller{\Delta}_{\alpha}}\left(\frac{x+x'}{2}\right)K^{*}_{0}(x,T;x'',0)K_{0}(x,T;x',0)\varphi^{*}_0(x'')\varphi_0(x')\nonumber\\
&\times \mathlarger{{\rm e}}^{-\displaystyle\-{\rm i}\phi\left(\frac{x+x''}{2},\omega,T\right)}\mathlarger{{\rm e}}^{\displaystyle{\rm i}\phi\left(\frac{x+x'}{2},\omega,T\right)}\mathlarger{\mathlarger{\int}}_{-\infty}^\infty{\rm d}X'\Phi_0^{*}\left(X'-g(\omega,T)\frac{x''-x'}{2}\right)\Phi_0(X')\,.
\end{align}
\end{widetext}
The overlap integral shows that for narrow~$\Phi_0(X)$, the values of~$x'$ and~$x''$ that will contribute to the decoherence functional are those for which~$x'\approx x''$. The expression~\eqref{decoherence-functional3} will be used throughout the rest of this work.

Next we recall an important sum rule for the decoherence functional.  We start with the property in Ref.~\cite{H1} that the sum of all the class operators is equal to the time evolution operator for the system
\begin{equation}\label{sum-clss-ops}
\sum_\alpha\widehat C_\alpha ={\rm e}^{-{\rm i}HT}\,.
\end{equation}
Inserting this expression into~\eqref{decoherence-functional} we obtain
\begin{equation}\label{property}
\sum_{\alpha,\alpha'}D(c_{\alpha},c_{\alpha'})=\Braket{\Psi_0|\Psi_0},
\end{equation} 
and in the case of decoherence it follows that
\begin{equation}\label{probability}
\sum_\alpha p_\alpha=\Braket{\Psi_0|\Psi_0},
\end{equation}
where
\begin{equation*}
p_\alpha=D(c_\alpha,c_\alpha)
\end{equation*}
is the probability that can by assigned to the alternative~$\alpha$. 
Clearly only when the initial state vector is normalizable do we get the sum of all probabilities to be unity. If the initial state vector is not normalizable then we obtain relative probabilities.

In Appendix~\ref{appendixA} we show that the property~\eqref{property} is satisfied by the exact expression~\eqref{decoherence-functional3} for the decoherence functional.

In the next section we will examine exact decoherence for the cases that at~$t=0$ 
(a) the particle is in a position eigenstate, and (b) the pointer is in a position eigenstate.


\section{Exact decoherence}
\label{exact decoherence}

The decoherence functional depends on the initial state of the system. In this section we consider what kind of initial states lead to exact decoherence. We will find that these states will not be normalizable and thus the probabilities that can be assigned will be relative.  This will follow from the general result~(\ref{probability}) regarding the sum of all the probabilities.


\subsection{ A sharp initial state  for the particle}

The initial wavefunction of the system is
\begin{equation}\label{initial-wavefunction1}
\Psi_0(x,X)=\delta(x-x_0)\Phi_0(X),
\end{equation}
where the initial pointer wavefunction~$\Phi_0(X)$ is a normalized wavepacket centered at~$X=0$ and~$\varphi_0(x)=\delta(x-x_0)$ is the sharp position initial wavefunction of the particle at~$x_0$.  

From~\eqref{branch-wavefunction3} the branch wavefunction can be readily obtained: 
\begin{align}\label{branch-wf}
\Psi_\alpha(x,X,T)=&\int_{-\infty}^\infty{\rm d}X'K_{TOT}(x,X,T;x_0,X',0)\nonumber\\
&\times
\Phi_0(X')
\end{align}
with\quad$2\bar x_\alpha-x-\delta<x_0<2\bar x_\alpha-x+\delta$, and~$\Psi_\alpha(x,X,T)=0$ otherwise.
 After inserting~\eqref{PropTOT1} and~\eqref{Phi(X,T)} in~\eqref{branch-wf} above we further obtain
\begin{align}\label{branch-wf.cases}
\Psi_\alpha=
\begin{cases}
~~K_0(x,T;x_0,0)\exp\left[{\rm i}\phi\left(\frac{x+x_0}{2},\omega,T\right)\right]
\\\
~\times
\Phi\big(X-s(x,x_0)\big),\quad{\rm with}\quad\frac{x+x_0}{2}\in\Delta_\alpha\\\\
~~0,\quad{\rm with}\quad\frac{x+x_0}{2}\notin\Delta_\alpha\,\,\,.
\end{cases}
\end{align}

\begin{figure}[h] 
	\psset{unit=1.5cm}
	\center
	\begin{pspicture}(-3,-1.8)(3,2.3)

	\psline[linewidth=.75pt](-2.5,0)(2.3,0)
	\psline{->,arrowlength=1, arrowsize=0.15,arrowinset=2}(2.2,0)(2.3,0)
	
	\rput[h](2.6,0){$\mathsmaller{\mathsmaller{x}}$}
	
	\psline[linewidth=.25pt] (-.75,-0.2)(-.75,0.2)
	\psline[linewidth=.25pt] (-2,-0.2)(-2,0.2)
	
	\psline[linecolor=blue,linestyle=dashed,linewidth=0.5pt] (-.75,0)(-.75,1.2)
	\psline[linecolor=blue,linestyle=dashed,linewidth=0.5pt] (-2,0.)(-2,1.5)

	\pscurve[linecolor=red,linewidth=2pt](-2,1.5)(-1.25,1.9)(-.75,1.2)
	
	\rput[bh](-2.1,-0.4){$\mathsmaller{\mathsmaller{2\bar x_\alpha-x_0-\delta}}$}
	\rput[bh](-.8,-0.4){$\mathsmaller{\mathsmaller{2\bar x_\alpha-x_0+\delta}}$}
	
	\psline[linewidth=.01]{->,arrowlength=1, arrowsize=0.15,arrowinset=2}(-2.3,1)(-2.3,1.7)
	\rput[vh](-2.7,1.4){$\mathsmaller{\mathsmaller{\Psi_\alpha(x)}}$}

	\psline[linecolor=blue,linestyle=dashed,linewidth=0.5pt] (.5,0.)(.5,1.2)
	\psline[linecolor=blue,linestyle=dashed,linewidth=0.5pt] (1.75,0)(1.75,1.5)
	
	\pscurve[linecolor=red,linewidth=2pt](.5,1.2)(1.25,1.)(1.75,1.5)
	
	\psline[linewidth=.25pt] (.5,-0.2)(.5,0.2)
	\psline[linewidth=.25pt] (1.75,-0.2)(1.75,0.2)
	
	\rput[bh](.47,-0.4){$\mathsmaller{\mathsmaller{2\bar x_{\alpha'}-x_0-\delta}}$}
	\rput[bh](1.72,-0.4){$\mathsmaller{\mathsmaller{2\bar x_{\alpha'}-x_0+\delta}}$}
	
	\psline[linewidth=.01]{->,arrowlength=1, arrowsize=0.15,arrowinset=2}(2,1)(2,1.7)
	\rput[vh](2.4,1.4){$\mathsmaller{\mathsmaller{\Psi_{\alpha'}(x)}}$}

	\end{pspicture}


    \par\vspace{-14ex}
\caption{ A schematic representation of two branch wavefunctions~$\Psi_\alpha$ and~$\Psi_{\alpha'}$ as a function of~$x$.  The pointer variable~$X$ is kept at a constant value in~$-\infty<X<\infty$. The~$\psi_\alpha(x,X,T)$ branch wavefunction vanishes for points such that~$(x+x_0)/2\notin\Delta_\alpha$ and similarly~$(x+x_0)/2\notin\Delta_{\alpha'}$ for~$\Psi_{\alpha'}$, with~$x_0$ being the initial position of the particle. There is no overlap between~$\Psi_\alpha$ and~$\Psi_{\alpha'}$ for~$\alpha'\not =\alpha$ and decoherence ensues.}
	
	\label{fig1}
	\end{figure}
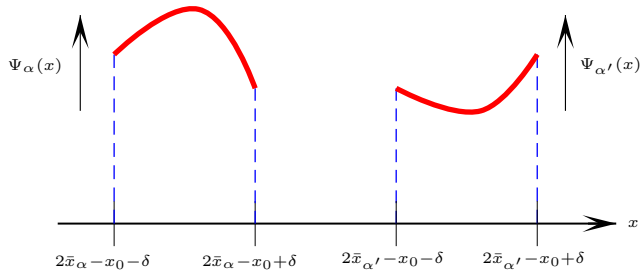

We can see from this expression that there will be exact decoherence because there is no overlap between the different branch wavefunctions.  This is illustrated in Figure~\ref{fig1} for two branch wavefunctions with~$\alpha\neq\alpha'$. 

Next we can add the branch wavefunctions for each of the coarse-grained alternatives in the set  
\[
C=\bigcup_{\alpha\in \mathbb{Z}}c_{\alpha}.
\]
That is,
\begin{equation}
\sum_\alpha\Psi_\alpha=\sum_\alpha\Braket{x,X|\widehat C_\alpha^{\mathit system}|\Psi_0}.
\end{equation}
 Inserting the general result from references~\cite{H1,H2}
\begin{equation}
\sum_\alpha\widehat C_\alpha^{\mathit system}={\rm e}^{-{\rm i}HT},
\end{equation}
where the Hamiltonian~$H$ of the system is given by~\eqref{H}, we obtain the result
\begin{equation}
\sum_\alpha\Psi_\alpha=\Psi(x,X,T),
\end{equation}
with 
\begin{equation}
\Psi(x,X,T)=\Braket{x,X|{\rm e}^{-{\rm i}HT}|\Psi_0}.
\end{equation}
Thus the sum of the branch wavefunctions, one for each alternative, is equal to the wavefunction of the system at time~$T$.   Decoherence implies the orthogonality of the different branch wavefunctions.

In the present case~$\Psi_\alpha(x,X,T)$ is the portion of the wavefunction~$\Psi(x,X,T)$ of the system truncated to the interval~$2\bar x_\alpha-x_0-\delta<x<2\bar x_\alpha-x_0+\delta$.  This is also obvious from the expressions~\eqref{class-op-matrix-elements1} and~\eqref{branch-wavefunction1} for the matrix elements of the class operators and the branch wavefunction respectively.

Next we will proceed to the explicit evaluation of the decoherence functional.
Inserting~\eqref{initial-wavefunction1} into the general expression~\eqref{decoherence-functional3} yields
\begin{align}\label{decoherence-functional4}
D(c_{\alpha},c_{\alpha'})&=\int_{-\infty}^\infty\!\!{\rm d}x\,\mathlarger{e}_{\mathsmaller{\Delta}_{\alpha'}}\left(\frac{x+x_0}{2}\right)\mathlarger{e}_{\mathsmaller{\Delta}_{\alpha}}\left(\frac{x+x_0}{2}\right)\nonumber\\
&\times
K^{*}_{0}(x,T;x_0,0)K_{0}(x,T;x_0,0).
\end{align}
Next, if
\begin{equation}
\frac{x+x_0}{2}\in\Delta_\alpha\quad{\rm then}\quad\frac{x+x_0}{2}\notin\Delta_{\alpha'},\quad\alpha\not =\alpha'
\end{equation}
then
\begin{equation}
D(c_{\alpha},c_{\alpha'})=0,\quad\alpha\not =\alpha'\,.
\end{equation}
Thus exact decoherence is obtained and probabilities can be assigned to the different alternatives.

 The probabilities follow from~\eqref{decoherence-functional4} by letting~$\alpha'=\alpha$. That is,
\begin{align}\label{probabilities1}
p_\alpha= &\int_{-\infty}^\infty{\rm d}x\,\mathlarger{e}_{\mathsmaller{\Delta}_{\alpha}}\left(\frac{x+x_0}{2}\right)\nonumber\\
\times&
K^{*}_{0}(x,T;x_0,0)K_{0}(x,T;x_0,0).
\end{align}
For quadratic potentials the propagator~$K_0$ takes the form
\begin{equation}\label{propagator-quadratic-pot}
K_0(x,T;x_0,0)=A(T){\rm e}^{{\rm i}S_{\mathit{cl}}(x,x_0)}
\end{equation}
with ~$S_{\mathit{cl}}$ being a classical action such as the action for a free harmonic oscillator.
Then from~\eqref{probabilities1} it follows that
\begin{equation}
p_\alpha=2\delta|A(T)|^2.
\end{equation}
For the driven oscillator we obtain
\begin{equation}\label{probabilities2}
p_\alpha=\frac{m\omega\delta}{\pi\sin\omega T}\,.
\end{equation}
This result also follows readily from
\begin{equation}
p_\alpha=\Braket{\Psi_\alpha|\Psi_\alpha}
\end{equation}
after inserting the expression~\eqref{branch-wf.cases} for the branch wavefunction and using
\begin{equation}
\int_{-\infty}^\infty\big|\Phi\big(X-s(x,x'),T\big)\big|^2{\rm d}X=1,
\end{equation}
which follows from~$\Braket{\Phi_0|\Phi_0}=1$ and unitary evolution.
These relative probabilities~\eqref{probabilities2} are the same for all alternatives and are indifferent to the the value of the initial position of the particle.  Also,
\begin{equation}
\sum_\alpha p_\alpha=\infty\,,
\end{equation}
which is consistent with the expression~\eqref{probability} for the sum over all probabilities when the initial state  for the system cannot be normalized.  Furthermore this result does not depend on the presence of a driving force, nor does the initial wavefunction for the pointer as long as it is normalized.

Next we will look at decoherence in the case of the particle by itself. That is, there is no coupling to a measuring apparatus.  


\subsubsection{Particle not coupled to the pointer}
In this case the particle does not not interact with a von Neumann apparatus.
The propagator for the particle is given by the result for a driven harmonic oscillator
\begin{align}
\Braket{x|{\rm e}^{-{\rm i}H_{\mathit{particle}}T}|x'}=&K_0(x,T;x',0)\nonumber\\
&\times
\exp\left[{\rm i}\phi\left(\frac{x+x'}{2},\omega,T \right)\right]
\end{align}
with
\begin{equation}
H_{\mathit{particle}}=\frac{p^2}{2m}+V(x,t),
\end{equation}
and~$V(x,t)$ is given by~\eqref{V driven SHO}.
The phase~$\phi$ is given in reference~\cite{FR-1} by Eq.(33). That is,
\begin{equation} \label{eq:Phi}
\phi =\frac{1}{\sin\omega T}\left[\left(\frac{x+x'}{2}\right)B_D-\frac{A_D}{m\omega}\right],
\end{equation}
with
\begin{align} \label{eq:A_D}
A_D(\omega,T)=\int_0^T\int_0^t&{\rm d}t\,{\rm d}sf_D(t)f_D(s)\nonumber\\
&\times\sin\omega (T-t)\sin\omega s\,,
\end{align}
and
\begin{equation} \label{eq:B_D}
B_D(\omega,T)=2\int_0^T{\rm d}t{f_D(t)\sin\omega t}.
\end{equation}
The matrix elements of the class operators are 
\begin{align}\label{class-op-matrix-elements2}
\Braket{x|\widehat C_{\alpha}|x'}=& \,\,\mathlarger{e}_{\mathsmaller{\Delta}_{\alpha}}\left(\frac{x+x'}{2}\right)\nonumber\\
&\times
\Braket{x|{\rm e}^{-{\rm i}H_{\mathit{particle}}T}|x'}.
\end{align}
The only change in the decoherent functional expression~\eqref{decoherence-functional3} is to set~$\Delta_{\mathit overlap}=1.$

When the initial state of the particle is described by a sharp wavefunction which is an eigenstate of position at~$x_0$ we obtain expression~\eqref{decoherence-functional4} which shows exact decoherence.

The probability for the different alternatives is again given by the result~\eqref{probabilities2}.  Thus, decoherence ensues whether the particle is coupled or not coupled to the apparatus.  All that matters is that the initial state of the particle corresponds to a sharp value of the position.

 
\subsection{A sharp initial state for the pointer}
The initial wavefunction of the system is
\begin{equation}\label{initial-wavefunction2}
\Psi_0(x,X)=\varphi_0(x)\delta(X)
\end{equation}
where the initial normalized particle wavefunction is~$\varphi_0(x)$, and~$\Phi_0(x)=\delta(X)$ is the  initial wavefunction  for the pointer.
The overlap integral~\eqref{pointer-overlap} for the pointer becomes
\begin{equation}\label{pointer-overlap1}
\Delta_{\mathit{overlap}}=\frac{2}{g(\omega,T)}\delta(x''-x').
\end{equation}
Introducing~\eqref{pointer-overlap1} into~\eqref{decoherence-functional3} we obtain
\begin{align}\label{decoherence-functional5}
D(c_{\alpha},c_{\alpha'})&=\frac{2}{g(\omega,T)}\int_{-\infty}^\infty {\rm d}x\int_{-\infty}^\infty {\rm d}x'|\varphi_0(x')|^2\nonumber\\
&\times
\mathlarger{e}_{\mathsmaller{\Delta}_{\alpha'}}\left(\frac{x+x'}{2}\right)\mathlarger{e}_{\mathsmaller{\Delta}_{\alpha}}\left(\frac{x+x'}{2}\right)\nonumber\\
&\times
|K_0(x,T,x',0)|^2\, .
\end{align}
This expression for the decoherence functional contains the factor~$\mathlarger{e}_{\mathsmaller{\Delta}_{\alpha'}}\left(\frac{x+x'}{2}\right)\mathlarger{e}_{\mathsmaller{\Delta}_{\alpha}}\left(\frac{x+x'}{2}\right)$
and therefore
\[
D(c_{\alpha'},c_\alpha)=0,\quad\alpha\not =\alpha'.
\]
Thus exact decoherence.

The probability for each alternative can be readily obtained from~\eqref{decoherence-functional5}
\begin{align*}
p_\alpha=&\frac{2}{g(\omega,T)}{\mathlarger\int}_{\!\!\!\!\!-\infty}^{\infty}{\rm d}x{\mathlarger\int}_{\!\!\!\!\!-\infty}^\infty{\rm d}x'\left[\mathlarger{e}_{\mathsmaller{\Delta}_{\alpha}}\left(\frac{x+x'}{2}\right)\right]^2\nonumber\\
&\times
|K_0(x,T;x',0)|^2|\varphi_0(x')|^2\,.
\end{align*}
For quadratic potentials~$K_0$ is of the form~\eqref{propagator-quadratic-pot}.
We can rewrite
\begin{align}
p_\alpha=&\frac{2}{g(\omega,T)}|A(T)|^2{\mathlarger\int}_{\!\!\!\!\!-\infty}^{\,\,\infty}{\rm d}x'|\varphi_0(x')|^2\nonumber\\
&\times
{\mathlarger\int}_{\!\!\!\!2\bar x_{\alpha}-x-\delta}^{2\bar x_{\alpha}-x+\delta}{\rm d}x\,\mathlarger{e}_{\mathsmaller{\Delta}_{\alpha}}\left(\frac{x+x'}{2}\right),
\end{align}
and with a normalized initial particle state we obtain
\begin{equation}\label{probabilities3}
p_\alpha=\frac{4\delta}{g(\omega,T)}|A(T)|^2 .
\end{equation}
For a driven oscillator the driving force does not affect the probabilities and we obtain
\begin{equation}\label{probabilities4}
p_\alpha=\frac{m\omega}{g(\omega,T)\pi\sin\omega T}2\delta\,.
\end{equation}
 In Ref.~\cite{FR-1} we obtained the expression~\eqref{shiftfunction} for the shift that the pointer of the apparatus experiences as a result of the finite time measurement. The coupling constant~$g$ in Ref.~\eqref{shiftfunction} is given by
  \begin{equation}\label{coupling const-g}
g(\omega, T)=\frac{B(\omega,T)}{T\sin\omega T},
\end{equation}
where
\begin{equation}
B(\omega,T)=2\int_0^T{\rm d}tf(t)\sin\omega t,
\end{equation}
 as shown in Ref.~\cite{FR-1},
and~$f(t)$ is the coupling function in the interaction Hamiltonian~\eqref{Hi}.

If~$g(\omega,T)\rightarrow 0$ then~$p_\alpha\rightarrow \infty$. This occurs because the pointer becomes decoupled from the particle. With~$\Braket{\Phi_0|\Phi_0}\rightarrow \infty$, then with vanishing coupling in~\eqref{pointer-overlap},~$\Delta_{\mathit{overlap}}\rightarrow\infty$.  Thus~\eqref{decoherence-functional3} yields~$D(\alpha,\alpha)\rightarrow\infty$.

The relative probabilities~\eqref{probabilities4} are the same for all alternatives and are indifferent to the initial wavefunction of the particle.  Also, the sum of the probabilities is consistent with~\eqref{probability} for an non-normalizable initial state for the system. 

\section{Approximate decoherence for narrow initial states.  Qualitative description}
\label{Section IV}
In this section we examine initial narrow states first for the pointer and then for the particle.
We will show qualitatively how they lead to approximate decoherence.  Detailed quantitative examples will be developed in  subsequent sections.


\subsection{Narrow initial pointer states}
 We begin with a normalized Gaussian pointer
\begin{equation}\label{gaussin pointer}
\Phi_0(X)=A{\rm e}^{- X^2/\ell^2},
\end{equation}
with
\begin{equation}\label{normalization}
A=\left(\frac{2}{\ell^2\pi}\right)^{1/4},
\end{equation}
and~$\ell\ll\delta$.
That is, the width of the Gaussian is very small compared to the width of the intervals~$\Delta_\alpha$ and the value of the pointer overlap integral~\eqref{pointer-overlap} is
\begin{equation}\label{gaussian-overlap}
\Delta_{\mathit{overlap}}={\exp}\left[-\frac{1}{2\ell^2} g^2(\omega,T)\left(\frac{x''-x'}{2}\right)^2\right].
\end{equation}
When this expression for~$\Delta_{\mathit{overlap}}$  is inserted into the exact expression~\eqref{decoherence-functional3} for the decoherence functional, it is clear that the main contributions to the integrals over~$x'$ and~$x''$  come from the region where the exponent in~\eqref{gaussian-overlap} is approximately unity. That is, 
\begin{equation}\label{inequalities}
-\frac{2\sqrt{ 2}}{g(\omega,T)}\ell\leqslant x''-x'\leqslant \frac{2\sqrt{ 2}}{g(\omega,T)}\ell.
\end{equation}
 Thus~$x'$ and~$x''$ are close to each other compared to~$\delta$.  For any value~$x$ in the integral over~$x$ in~(\ref{decoherence-functional3})
 \begin{equation}\label{Delta-Delta}
\frac{x+x'}{2}\in\Delta_\alpha\quad{\rm and}\quad\frac{x+x''}{2}\in\Delta_{\alpha'}
\end{equation}
and the main contribution to the decoherence functional appears when~$(x+x')/2$ and~$(x+x'')/2$ are in contiguous intervals.  Thus we can set~$\alpha'=\alpha+1$ because
\begin{equation*}
|x''-x'|<\ell\ll\delta. 
\end{equation*}
The expressions~\eqref{Delta-Delta} can be rewritten
\begin{align}\label{inequalities1}
&\bar x_\alpha-\frac{\delta}{2}<\frac{x+x'}{2}\leqslant\bar x_\alpha+\frac{\delta}{2}\nonumber\\
&\bar x_{\alpha+1}-\frac{\delta}{2}<\frac{x+x''}{2}\leqslant\bar x_{\alpha+1}+\frac{\delta}{2}
\end{align}
with
\begin{equation}\label{barx_alpha+1=barx_alpha+delta}
\bar x_{\alpha +1}-\bar x_\alpha =\delta.
\end{equation}
 Further from~\eqref{inequalities1} and~\eqref{barx_alpha+1=barx_alpha+delta}
\begin{align}\label{inequalities2}
&2\bar x_\alpha-\delta-x<x'\leqslant 2\bar x_\alpha+\delta-x\nonumber\\
&2\bar x_\alpha+\delta-x<x''\leqslant 2\bar x_{\alpha+1}+\delta-x\,.
\end{align}
For every value~$-\infty<x<\infty$ , the inequalities~\eqref{inequalities} and~(\ref{inequalities2}) set the significant integration ranges for~$x'$ and~$x''$ in ~\eqref{decoherence-functional3}:
\begin{align}\label{inequalities3}
&2\bar x_\alpha-x+\delta-\frac{\ell}{g}\sqrt{2}\leqslant x'\leqslant 2\bar x_\alpha-x+\delta\nonumber\\
&2\bar x_\alpha-x+\delta<x''\leqslant 2\bar x_\alpha-x+\delta+\frac{\ell}{g}\sqrt{2}\,.
\end{align}
Thus these integrations in~(\ref{decoherence-functional3}) are carried out over a very small interval of size~$\ell\sqrt{2}/g$.  

\begin{figure}[h] 
	\psset{unit=1cm}
	\begin{pspicture}(-2,-1.5)(2,1.5)
	%
	\psline[linewidth=1pt](-3.5,0)(3.5,0)
	\psline[linestyle=dashed,linewidth=.5pt,linecolor=blue] (0,-1)(0,1)
	\psline[linewidth=0.5pt] (1.2,-.25)(1.2,.25)
	\psline[linewidth=0.5pt] (-1.2,-.25)(-1.2,.25)
	\psline[linewidth=0.5pt] (3.5,-.25)(3.5,.25)
	\psline[linewidth=0.5pt] (-3.5,-.25)(-3.5,.25)

	\rput[bh](-1.2,-0.8){$\mathsmaller{x_b-\frac{\ell\sqrt 2}{2g}}$}
	\rput[bh](1.2,-0.8){$\mathsmaller{x_b+\frac{\ell\sqrt 2}{2g}}$}
	\rput[bv](-0.4,0.2){$\mathsmaller{\frac{x+x'}{2}}$}
	\rput[bv](0.5,0.2){$\mathsmaller{\frac{x''+x}{2}}$}

	\psline[linewidth=0.5pt] (.5,-.10)(.5,.10)
	\psline[linewidth=0.5pt] (-.4,-.10)(-.4,.10)

	\rput[bh](3.5,0.4){$\mathsmaller{\bar x_{\alpha+1}+\frac{\delta}{2}}$}
	\rput[bh](-3.5,0.4){$\mathsmaller{\bar x_{\alpha}-\frac{\delta}{2}}$}

	\rput[bc](0.05,-1.2){$\mathsmaller{x_b}$}
  
	\end{pspicture}
	\caption{The relative position of~$\mathsmaller{\bar x'=(x+x')/2}$ and~$\mathsmaller{\bar x''=(x+x'')/2}$
  	in the contiguous intervals~$\Delta_\alpha$ and~$\Delta_{\alpha+1}$. The vertical dashed 	line is the common boundary at~$\mathsmaller{x_b=\bar x_\alpha+\frac{\delta}{2}=\bar x_{\alpha+1}-\frac	{\delta}{2}}$. The length of each interval is~$\delta$. The length~$\ell$ is 	much smaller than~$\delta$; and~$\mathsmaller{\frac{\ell\sqrt 2}{g}}$ is the size of the 	region of integration over~$x'$ and~$x''$ }
	\label{fig2}
	\end{figure}
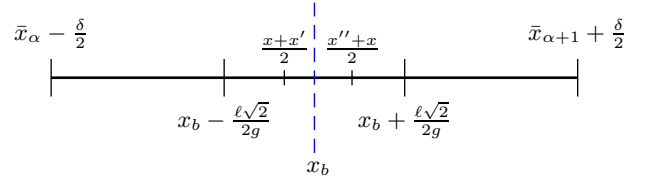

Figure~\ref{fig2} shows the relative position of~$\bar x'=(x+x')/2$ and~$\bar x''=(x+x'')/2$ and the relative size of the intervals.  The regions with a significant contribution to the~$x'$ and~$x''$ integrations are also shown.

To lowest order in~$\ell$ we can set~$x'=2\bar x_\alpha-x+\delta$ in the integration over~$x'$, and similarly in the integration over~$x''$.  
Setting~$x'\approx x''$ we can approximate the decoherence functional
\begin{align}\label{decoherence-functional6}
D(c_{\alpha},c_{\alpha+1})&\approx\int_{-\infty}^\infty{\rm d}x\int_{2\bar x_\alpha-x+\delta}^{2\bar x-x+\delta+\frac{\ell}{g}\sqrt 2}{\rm d}x''\nonumber\\
&\times
\int_{2\bar x_\alpha-x+\delta-\frac{\ell}{g}\sqrt 2}^{2\bar x_\alpha-x+\delta}{\rm d}x'|K_0(x,T;x',0)|^2\nonumber\\
&\times
\varphi_0^{*}(x')\varphi_0(x')\,.
\end{align}
Next we set 
\begin{equation}
x''\approx x'=2\bar x_\alpha-x+\delta
\end{equation}
in the integrants corresponding to the integrations over~$x'$ and~$x''$ in~\eqref{decoherence-functional6},
and from Ref.~\cite{Feynman}
\begin{equation}
|K_0(x,T;x',0)|^2=\frac{m\omega}{2\pi\sin\omega T}
\end{equation}
 we obtain
\begin{align}
D(c_{\alpha},c_{\alpha+1})&\approx\frac{m\omega\ell^2}{\pi g^2(\omega,T)\sin\omega T}\nonumber\\
&\times
\int_{-\infty}^\infty{\rm d}x|\varphi_0(2\bar x_\alpha-x+\delta)|^2,
\end{align}
and for a normalized particle wavefunction~$\varphi_0(x)$ follows the result
\begin{equation}\label{pointer-decoherene}
D(c_{\alpha},c_{\alpha+1})\approx\frac{m\omega \ell^2}{\pi g^2(\omega,T)\sin\omega T}.
\end{equation}
Since~$\ell<<\delta$, the decoherence functional is of the second order in smallness in~$\mathsmaller{\frac{\ell}{\delta}}$.  

 For~$\alpha'=\alpha\pm2$ the decoherence functional is even smaller.
 That is
\begin{equation}
D(c_\alpha,c_{\alpha'})\approx 0
\end{equation}
where~$\alpha'=\alpha+n\quad{\rm with}\quad n=\pm 2,\pm 3,\pm 4,\dots$\\

Thus there is approximate decoherence for narrow initial pointer states and approximate probabilities can be assigned to the different coarse-grained alternatives~$\alpha$.

\pagebreak


 \subsection{Narrow initial particle states}
For the sake of simplicity we will assume that the wavefunction of the initial state of the particle is described by a Gaussian centered at~$x_0$ with a narrow spread~$\sigma\ll\delta$
 \begin{equation}\label{gaussian}
 \varphi_0(x)=A{\rm e}^{-(x-x_0)^2/\sigma^2}.
 \end{equation} 
  Next in the exact expression~\eqref{decoherence-functional3} for the decoherence functional the values of the variables~$x'$ and~$x''$ can vary around~$x_0$ within a very small range~$(x_0-\sigma,\,x_0+\sigma)$, where~$\sigma\ll\delta$. Thus their significant contributions to the integrations in~\eqref{decoherence-functional3} are determined by the ranges~$x_0-\sigma<x'<x_0+\sigma$ and~$x_0-\sigma<x''<x_0+\sigma$. The value of~$x_0$ is fixed, and it can be any real number.
  
 From~\eqref{decoherence-functional3} and after inserting the initial wavefunction~\eqref{gaussian} we can write the inequality
\begin{align}\label{abs-decoherent-functional1}
|D(c_\alpha,c_{\alpha'})|&\leqslant\mathlarger{\int}_{-\infty}^\infty {\rm d}x
\mathlarger{\int}_{2\bar x_{\alpha'}-x-\delta}^{2\bar x_{\alpha'}-x+\delta}{\rm d}x''A^{*}{\rm e}^{-(x''-x_0)^2/\sigma^2}\nonumber\\
&\times
\mathlarger{\int}_{2\bar x_{\alpha}-x-\delta}^{2\bar x_{\alpha}-x+\delta}{\rm d}x'A{\rm e}^{-(x'-x_0)^2/\sigma^2}|K_0|^2\nonumber\\
&\times
\mathlarger{\Bigg|}\mathlarger{\mathlarger{\int}}_{-\infty}^\infty{\rm d}X'\Phi_0^{*}\left(X'-g(\omega,T)\frac{x''-x'}{2}\right)\nonumber\\
&\times
\Phi_0(X')\mathlarger{\Bigg|}.
 \end{align}
 
 In the lowest order in~$|x''-x'|$ we can approximate the overlap integral~\eqref{pointer-overlap} for the pointer by setting~$x''\approx x'$. That is, for a normalized initial state of the pointer
\begin{align}
\mathlarger{\mathlarger{\int}}_{-\infty}^\infty&{\rm d}X'\Phi_0^{*}\left(X'-g(\omega,T)\frac{x''-x'}{2}\right)\nonumber\\
&\times
\Phi_0(X')= 1+O(|x''-x'|).
\end{align}
Since only the intervals~$\Delta_\alpha$ and~$\Delta_{\alpha+1}$ give a relevant 
\pagebreak
contribution to~\eqref{abs-decoherent-functional1}, we can thus write the inequality
\begin{align}\label{abs-decoherent-functional2}
|D(c_\alpha,c_{\alpha+1})|\leqslant&\frac{m\omega}{2\pi \sin\omega T}|A|^2\mathlarger{\int}_{-\infty}^\infty {\rm d}x\nonumber\\
&\times
\mathlarger{\int}_{2\bar x_\alpha-x+\delta}^{2\bar x_{\alpha}-x+3\delta}{\rm d}x''{\rm e}^{-(x''x_0)^2/\sigma^2}\nonumber\\
&\times
\mathlarger{\int}_{2\bar x_{\alpha}-x-\delta}^{2\bar x_{\alpha}-x+\delta}{\rm d}x'{\rm e}^{-(x'-x_0)^2/\sigma^2},
\end{align}
where in~\eqref{abs-decoherent-functional1} we have inserted
\begin{equation}
|K_0|^2=\frac{m\omega}{2\pi \sin\omega T}
\end{equation}
and have made use of the relation
\begin{equation}
\bar x_{\alpha+1}=\bar x_\alpha+\delta\,.
\end{equation}
 We have 
\begin{equation}\label{belonging}
\frac{x+x'}{2}\in\Delta_\alpha\quad{\rm and}\quad\frac{x+x''}{2}\in\Delta_{\alpha+1}\,,
\end{equation}
then with~$x'\lesssim x''$ it follows that
\begin{equation}\label{inequality1}
2\bar x_\alpha-x''+\delta\lesssim x\lesssim 2\bar x_\alpha-x'+\delta.
\end{equation}
 Figure~\ref{fig3} assumes that~$(x+x_0)/2$ is located at the common boundary of~$\Delta_\alpha$ and~$\Delta_{\alpha+1}$.  However, in general, we can have the situations\,\,$ x'<x_0<x'',\,\,
 x'<x''<x_0,~{\rm and}~x_0<x'<x''$, with the only condition that~$(x+x')/2\in\Delta_\alpha$ and~$(x+x'')/2\in\Delta_{\alpha+1}$.  Within the confines of the lowest order approximation\quad$x'\approx x''\approx x_0$ the only important condition is~$x'<x''$.
The relevant values of~$x'$ and $x''$ in the integrations in~\eqref{abs-decoherent-functional2} are within~$(x_0-\sigma,x_0+\sigma)$ with~$x_0\approx 2\bar x_\alpha-x+\delta$.  In addition, since  the significant ranges for~$x'$ and~$x''$ are
\begin{align}\label{inequalities3}
2\bar x_\alpha-x+\delta-\sigma\lesssim x'\lesssim 2\bar x_\alpha-x+\delta\nonumber\\
2\bar x_\alpha-x+\delta\lesssim x''\lesssim 2\bar x_\alpha-x+\delta+\sigma,
\end{align}
we can combine the inequalities~\eqref{inequality1} and~\eqref{inequalities3} together with the lowest order approximation~$x'\approx x''\approx x_0$ and deduce the effective range in~\eqref{abs-decoherent-functional2} for the integration over the variable~$x$
\begin{equation}\label{inequality3}
2\bar x_\alpha-x_0+\delta-\sigma\lesssim x \lesssim 2\bar x_\alpha-x_0+\delta+\sigma.
\end{equation}
Figure~\ref{fig4} illustrates the effective ranges of the different integrations in~\eqref{abs-decoherent-functional2} for the variables~$x'$ and~$x''$, and with the position of~$x_0$ on the common boundary of  the two contiguous regions~$\Delta_\alpha$ and~$\Delta_{\alpha+1}$.

 We can expand the exponentials in the integrands in~\eqref{abs-decoherent-functional2} in powers of~$(x'-x_0)^2$ and~$(x''-x_0)^2$ respectively.  That is,
\begin{align}\label{abs-decoherent-functional3}
|D(c_\alpha,&c_{\alpha+1})|\leqslant\frac{m\omega}{2\pi \sin\omega T}|A|^2\mathlarger{\int}_{\bar x_\alpha-x_0+\delta-\sigma}^{2\bar x_\alpha-x_0+\delta+\sigma} {\rm d}x\nonumber\\
&\times
\mathlarger{\int}_{2\bar x_\alpha-x+\delta}^{2\bar x_\alpha-x+\delta+\sigma}{\rm d}x''\big[1+O((x''-x_0)^2\big]\nonumber\\
&\times
\mathlarger{\int}_{2\bar x_{\alpha}-x+\delta-\sigma}^{2\bar x_\alpha-x+\delta}{\rm d}x'\big[1+O((x'-x_0)^2\big].
\end{align}
Thus after keeping the lowest order terms in the expansions we obtain the inequality for the decoherence functional
\begin{equation}\label{lowest order correction}
|D(c_\alpha,c_{\alpha+1})|\leqslant\frac{m\omega}{\pi \sin\omega T}|A|^2\sigma^3.
\end{equation}

If the particle initial wavefunction~$\varphi_0(x)$ is normalized we set~$|A|^2=\sqrt{\frac{2}{\pi}}\frac{1}{\sigma},$
and
\begin{equation}
|D(c_\alpha,c_{\alpha+1})|\leqslant\frac{m\omega}{\pi \sin\omega T}\sqrt{\frac{2}{\pi}}\sigma^2.
\end{equation}
If instead
\begin{equation}\label{phi->delta(x-x_0)}
\varphi_0(x)\xrightarrow[\sigma \rightarrow 0] {}\delta(x-x_0)
\end{equation}
then we set~$|A|^2=\frac{1}{\pi\sigma^2}$,
and
\begin{equation}
|D(c_\alpha,c_{\alpha+1})|\leqslant\frac{m\omega}{\pi^2 \sin\omega T}\sigma\,.
\end{equation}
Thus in either case we obtain approximate decoherence for narrow initial particle states with the strongest decoherence corresponding to the case of a normalized Gaussian for the initial state of the particle. 

In the next section we will carry out a detailed quantitative study of the approximate decoherence and approximate probabilities for a narrow Gaussian initial particle state and a Gaussian pointer and alternatively, a Gaussian initial particle state and a narrow and Gaussian pointer.



	 \begin{figure}[h] 

	 \psset{unit=.6} 

	\def\Gaussian{8*2.71828^(-2*x^2)} 
	\begin{pspicture}(-1.5,-6)(1.5,8)
	\psplot[linecolor=red,linewidth=2.pt,algebraic=true,plotpoints=100]{-2.5}{2.5}{\Gaussian} 

	\psline[linewidth=.01]{->,arrowlength=1, arrowsize=0.2,arrowinset=2}(-4,0)(4,0)

	\psline[linestyle=dashed,linecolor=blue,linewidth=.25pt] (0,8)(0,0)
	\psline[linewidth=.25pt] (0,-0.2)(0,0.2)

	\rput[vh](1.95,-0.6){$\mathsmaller{x_{\mathsmaller 0}\,=\,2\bar x_{\alpha}-x+\delta}$}

	\psline[linewidth=.25pt] (.55,-0.1)(.55,0.2)
	\rput[bv](.70,0.3){$\mathsmaller{x''}$}

	\psline[linewidth=.25pt] (-.6,-0.1)(-.6,0.2)
	\rput[bv](-.5,0.3){$\mathsmaller{x'}$}

	\psline[linewidth=.01]{<->,arrowlength=1, arrowsize=0.2,arrowinset=2}(-.55,3)(.55,3)
	\rput[vh](0,3.5){$\mathsmaller{2\,\sigma}$}


	\psline[linewidth=.01]{->,arrowlength=1, arrowsize=0.2,arrowinset=2}(-2,4)(-2,6)
	\rput[vh](-2.9,5.2){$\varphi_{\mathsmaller 0}(x)$}

	\rput[vh](4.5,0){$\mathsmaller{x}$}

	\end{pspicture}

\par\vspace{-22ex}
\caption{A narrow Gaussian centered  at~$x_0$ and half-width~$\sigma<<\delta$ describes the initial wavefunction of the particle. The case with~$x'<x_0<x''$ is illustrated with~$(x+x')/2\in\Delta_\alpha$ and~$(x+x'')/2\in\Delta_{\alpha+1}$.  The value of~$x_0=2\bar x_\alpha-x+\delta$ is such that~$(x_0+x)/2$ is located at the common boundary of~$\Delta_\alpha$ and~$\Delta_{\alpha+1}$. The values of~$x'$,~$x''$~and $x_0$ are very close to each other.}
\label{fig3}
\end{figure}
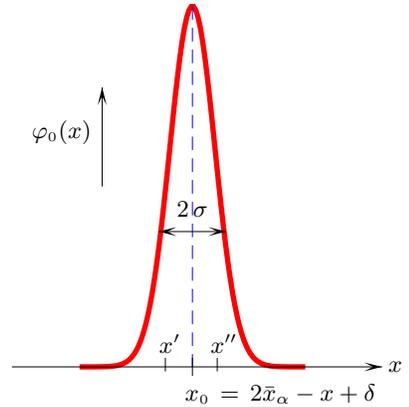

 	\begin{figure}[h] 
	\psset{unit=.9cm}
	\begin{pspicture}(-2,-1.5)(2,1.5)
	%
	\psline[linewidth=1pt](-3.5,0)(3.5,0)
	\psline[linestyle=dashed,linewidth=.5pt,linecolor=blue] (0,-1)(0,1)
	\psline[linewidth=0.5pt] (1.2,-.25)(1.2,.25)
	\psline[linewidth=0.5pt] (-1.2,-.25)(-1.2,.25)
	\psline[linewidth=0.5pt] (3.5,-.25)(3.5,.25)
	\psline[linewidth=0.5pt] (-3.5,-.25)(-3.5,.25)

	\rput[bh](-.9,0.4){$\mathsmaller{x_0-\sigma}$}
	\rput[bh](1.5,0.4){$\mathsmaller{x_0+\sigma}$}
	\rput[bv](-0.4,-0.5){$\mathsmaller{x'}$}
	\rput[bv](0.8,-0.5){$\mathsmaller{x''}$}

	\psline[linewidth=0.5pt] (.7,-.10)(.7,.10)
	\psline[linewidth=0.5pt] (-.4,-.10)(-.4,.10)

	\rput[bh](3.5,0.4){$\mathsmaller{2\bar x_{\alpha+1}-x+\delta}$}
	\rput[bh](-3.5,0.4){$\mathsmaller{2\bar x_\alpha-x-\delta}$}

	\rput[bc](0.05,1.2){$\mathsmaller{x_0}$}
	\rput[bc](0.15,-1.2){$\mathsmaller{2\bar x_\alpha-x+\delta\,=\,2\bar x_{\alpha+1}-x-	\delta}$}
  
	\end{pspicture}

\caption{The relative position of~$\bar x'=(x+x,)/2$ and~$\bar x''=(x+x'')/2$ in the contiguous intervals~$\Delta_\alpha$ and~$\Delta_{\alpha+1}$. The vertical dashed line is the common boundary at~$x_0$. The length of each interval is~$\delta$. The length~$\sigma$ is much smaller than~$\delta$, and~$(x_0-\sigma,2\bar x_\alpha -x+\delta)$ is the size of the region of integration over~$x'$. Furthermore~$(2\bar x_\alpha -x+\delta,2\bar x_\alpha -x+\delta+\sigma)$ is the effective region of integration of~$x''$ in the expression~(\ref{decoherence-functional3}) for the decoherence functional.}
\label{fig4}
\end{figure}
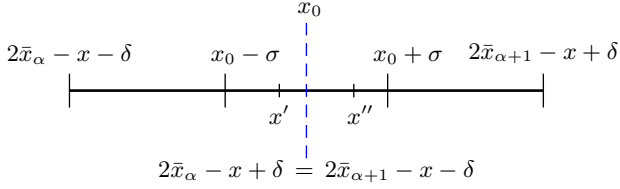
\pagebreak


\section{Narrow initial particle state.  Quantitative example}
\label{Section V} 
In this section we will assume a narrow Gaussian initial state for the particle interacting with a Gaussian pointer. At~$t=0$ the wavefunction of the particle is described by a Gaussian of half-width~$\sigma$
\begin{equation}
\varphi_0(x)=A{\rm e}^{-(x-x_0)^2/\sigma^2}\,,
\end{equation}
The pointer is described by a normalized Gaussian of half-width~$\ell$ as shown by~\eqref{gaussin pointer} and \eqref{normalization}.

The value of the pointer overlap integral~$\Delta_{\mathit overlap}$ is given by~\eqref{gaussian-overlap}.
For a narrow initial state of the particle we can write
\begin{equation}\label{x''-x'}
|x''-x'|\lesssim\sigma\ll\delta
\end{equation}
and
\begin{equation}
\frac{g^2}{8\ell^2} (x''-x')^2\ll 1,
\end{equation}
or from~\eqref{x''-x'}
\begin{equation}\label{condition}
\kappa^2\sigma^2\ll1,
\end{equation}
with
\begin{equation}\label{kappa}
\kappa^2=\frac{g^2(\omega,T)}{8\ell^2}\,.
\end{equation}

In the next two subsections we will evaluate expressions for the decoherence functional and assigned approximate probabilities.
\subsection{Decoherence}\label{narrow particle decoherence}
We can expand the result for the overlap integral~\eqref{gaussian-overlap} in powers of the exponent. Inserting this expansion into~\eqref{abs-decoherent-functional1} we obtain 
\begin{align}\label{|D|-expansion}
|D(c_\alpha,c_{\alpha+1})|\le &D^{(0)}(c_\alpha,c_{\alpha+1})+D^{(1)}(c_\alpha,c_{\alpha+1})\nonumber\\
&
+O(\bar\alpha^4),
\end{align}
where
\begin{align}\label{|D|^(0)}
D^{(0)}(c_\alpha,c_{\alpha+1})=&\frac{m\omega}{2\pi \sin\omega T}|A|^2\mathlarger{\int}_{-\infty}^\infty {\rm d}x\nonumber\\
&\times
\mathlarger{\int}_{2\bar x_\alpha-x+\delta}^{2\bar x_{\alpha}-x+3\delta}{\rm d}x''{\rm e}^{-(x''-x_0)^2/\sigma^2}\nonumber\\
&\times
\mathlarger{\int}_{2\bar x_{\alpha}-x-\delta}^{2\bar x_{\alpha}-x+\delta}{\rm d}x'{\rm e}^{-(x'-x_0)^2/\sigma^2},
\end{align}
and
\begin{align}\label{|D|^(2)}
D^{(1)}(c_\alpha,c_{\alpha+1})=&-\kappa^2\frac{m\omega}{2\pi \sin\omega T}|A|^2\mathlarger{\int}_{-\infty}^\infty {\rm d}x\,(x''-x')^2\nonumber\\
&\times
\mathlarger{\int}_{2\bar x_\alpha-x+\delta}^{2\bar x_{\alpha}-x+3\delta}{\rm d}x''{\rm e}^{-(x''-x_0)^2/\sigma^2}\nonumber\\
&\times
\mathlarger{\int}_{2\bar x_{\alpha}-x-\delta}^{2\bar x_{\alpha}-x+\delta}{\rm d}x'{\rm e}^{-(x'-x_0)^2/\sigma^2}.
\end{align}

\begin{figure}
	\centerline{\epsfxsize=3.5in\epsfbox{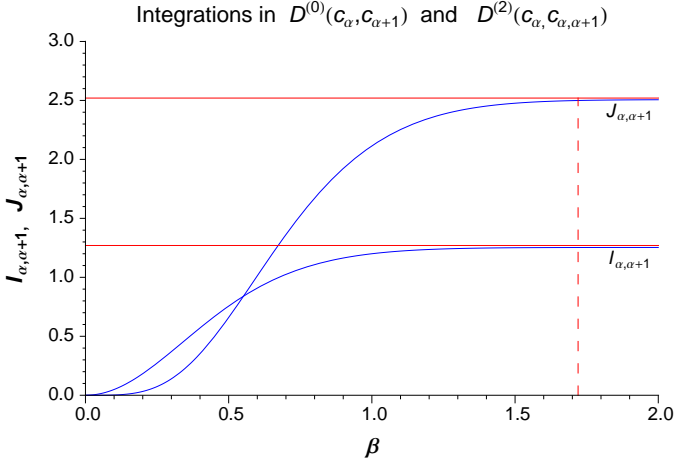}}
	\caption{The result of the integrations in~\eqref{|D|^(0)} and~\eqref{|D|^(2)} for a narrow initial state for the particle.  The solid red lines are 	the asymptotes at~$I_{\alpha,\alpha+1}=\sqrt{\pi/2}$ and~$J_{\alpha,\alpha+1}=\sqrt{2\pi}$.  The 	vertical red dotted line is at~$ \beta=1.72$ for which~$I_{\alpha,\alpha+1}=1.25$ and~$J_{\alpha,	\alpha+1}=2.49$.}
	 \label{fig5}
	\end{figure}

The integrations for the lowest order term are evaluated in Appendix ~\ref{AppendixB}.  Here we quote the result:
\begin{equation}\label{D^(0)-2}
D^{(0)}(c_\alpha,c_{\alpha+1})=|K_0|^2|A|^2\sigma^3I_{\alpha,\alpha+1},
\end{equation}
where
\begin{align}\label{I_a,a+1}
I_{\alpha,\alpha+1}&=\frac{1}{2}\sqrt{2\pi}-\sqrt{2\pi}\,{\rm e}^{-2\beta^2}+\frac{1}{2}\sqrt{2\pi}\,{\rm e}^{-8\beta^2}\nonumber\\
&
\!\!-2\pi\beta\,\erf\,\Big(\sqrt{2}\beta\Big)
\!+2\pi\beta\,\erf\,\Big(2\sqrt{2}\beta\Big)
\end{align}
with
\begin{equation*}
\beta=\frac{\delta}{\sigma}.
\end{equation*}
For a narrow initial state of the particle and letting~$\beta\rightarrow\infty$, we obtain the asymptotic behavior for the expression~\eqref{I_a,a+1}: 
\begin{equation}
I_{\alpha,\alpha+1}\rightarrow\frac{1}{2}\sqrt{2\pi}\,.  
\end{equation}
Fig~\ref{fig5} shows that this asymptotic value of~$I_{\alpha,\alpha+1}$ is reached at around~$\beta\gtrsim 1.6$. 

The integrations for the next higher order term are also found in Appendix B.  The result is
\begin{equation}\label{D^(2)-2}
D^{(1)}(c_\alpha,c_{\alpha+1})=-\kappa^2\sigma^5|K_0|^2|A|^2J_{\alpha,\alpha+1},
\end{equation}
\pagebreak
where
\begin{align}\label{J_a,a+1}
J_{\alpha,\alpha+1}&=\sqrt{2\pi}-2\sqrt{2\pi}\,{\rm e}^{-2\beta^2}+2\sqrt{2\pi}\,{\rm e}^{-8\beta^2}\nonumber\\
& 
\!\!-2\pi\beta\,\erf\,\Big(\sqrt{2}\beta\Big)
\!+2\pi\beta\,\erf\,\Big(2\sqrt{2}\beta\Big).
\end{align}
We finally obtain
\begin{align}
\!\!\!\!\!\!|D(c_\alpha,c_{\alpha+1)}|&\le|K_0|^2|A|^2\sigma^3\nonumber\\
&\times
\left(I_{\alpha,\alpha+1}-\kappa^2\sigma^2J_{\alpha,\alpha+1}\right).
\end{align}
For a narrow initial state of the particle~$\beta\rightarrow\infty$, and thus we obtain the asymptotic value
\begin{equation}
J_{\alpha,\alpha+1}\rightarrow\sqrt{2\pi}.  
\end{equation}

Fig~\ref{fig5} shows that this asymptotic value is reached at around
~$\beta=1.72$. 
Since~$\sigma<<\delta$ we can use the asymptotic values of~\eqref{D^(0)-2} and~\eqref{D^(2)-2} to obtain the result
\begin{align}\label{higher order corrections}
|D(c_\alpha,c_{\alpha+1})|&\le\sqrt{\frac{\pi}{2}}|K_0|^2|A|^2\sigma^3\nonumber\\
&\times
\Big(1-2\sigma^2\kappa^2+\dots\Big)
\end{align}
with~$\beta\gtrsim1.72$.  We can compare the result~\eqref{higher order corrections} with the result~\eqref{lowest order correction} obtained using semi-qualitative arguments for the estimation of the lowest order term in the expansion in powers of~$\sigma/\delta$ .

For a particle whose initial state is such that the limit~$\sigma\rightarrow 0$ is given by~\eqref{phi->delta(x-x_0)}, then
\begin{equation}
|A|^2=\frac{1}{\pi\sigma^2}
\end{equation}
and
\begin{align}
|D&(c_\alpha,c_{\alpha+1})|\le\frac{m\omega}{2\pi \sin\omega T}\frac{1}{\sqrt{2\pi}}\sigma\nonumber\\
&\times
\bigg\{1-\frac{1}{4}g^2(\omega,T)\frac{\sigma^2}{\ell^2}
+O\left[\left(\frac{\sigma}{\ell}\right)^4\right]\bigg\}.
\end{align}
For a normalized initial state
\begin{equation}
|A|^2=\left(\frac{2}{\pi\sigma^2}\right)^{1/2}
\end{equation}
and 
\begin{align}
|D&(c_\alpha,c_{\alpha+1})|\le\frac{m\omega}{2\pi \sin\omega T}\sigma^2\nonumber\\
&\times
\bigg\{1-\frac{1}{4}g^2(\omega,T)\frac{\sigma^2}{\ell^2}
+O\left[\left(\frac{\sigma}{\ell}\right)^4\right]\bigg\}.
\end{align}For both cases we need the condition~\eqref{condition}
\begin{equation}
\frac{\sigma}{\ell}\ll\frac{\sqrt{2}}{g(\omega,T)}<\frac{2}{g(\omega,T)}.
\end{equation}
\pagebreak
Decoherence is stronger as~$\sigma\rightarrow 0$, when the initial state of the particle is normalized.

Next we evaluate the approximate probabilities that can be assigned to the different coarse-grained alternatives in this example.
\subsection{Probabilities}\label{narrow-particle-prob}
 The approximate probabilities are defined according to
\begin{equation*}
p_\alpha\approx|D(c_\alpha,c_\alpha)|,
\end{equation*}
and the absolute value of the decoherent functional~$D({c_\alpha,c_\alpha})$ is given by
\begin{align}\label{abs-probabilities}
|D(c_\alpha,c_{\alpha})|&\leqslant|K_0|^2|A|^2\mathlarger{\int}_{-\infty}^\infty {\rm d}x\nonumber\\
&\times
\mathlarger{\int}_{2\bar x_\alpha-x+\delta}^{2\bar x_{\alpha}-x+\delta}{\rm d}x''{\rm e}^{-(x''x_0)^2/\sigma^2}\nonumber\\
&\times
\mathlarger{\int}_{2\bar x_{\alpha}-x-\delta}^{2\bar x_{\alpha}-x+\delta}{\rm d}x'{\rm e}^{-(x'-x_0)^2/\sigma^2}\nonumber\\
&\times
\Delta_{\it overlap}.
\end{align}
The general expression for~$\Delta_{\it overlap}$ is given in~\eqref{pointer-overlap}.
 After expanding~$\Delta_{\mathit overlap}$ in powers of~$\kappa$ we obtain
\begin{align}\label{approxprob}
p_{\alpha}\lesssim&\,|K_0|^2|A|^2|\mathlarger{\int}_{-\infty}^\infty {\rm d}x\nonumber\\
&\times
\mathlarger{\int}_{2\bar x_\alpha-x+\delta}^{2\bar x_{\alpha}-x+\delta}{\rm d}x''{\rm e}^{-(x''-x_0)^2/\sigma^2}\nonumber\\
&\times
\mathlarger{\int}_{2\bar x_{\alpha}-x-\delta}^{2\bar x_{\alpha}-x+\delta}{\rm d}x'{\rm e}^{-(x'-x_0)^2/\sigma^2}\nonumber\\
&\times
\left[1-\frac{g^2}{8\ell^2}(x''-x')^2+\dots\right]\,.
\end{align}
The detail of these integrations is shown in Appendix~\ref{AppendixB}. The result is expressed as a series expansion in powers of~$\sigma/\ell$                     
\begin{align}\label{prob expand}
p_\alpha\lesssim2&\pi|K_0|^2|A|^2\sigma^2\delta\nonumber\\
&
\bigg\{1-\frac{1}{8}g^2(\omega,T)\frac{\sigma^2}{\ell^2}+O\left[\left(\frac{\sigma}{\ell}\right)^4\right]\bigg\}\,.
\end{align}
If the initial state of the particle is such that 
\begin{equation*}
\varphi_0(x)\xrightarrow[\sigma \rightarrow 0] {}\delta(x-x_0)
\end{equation*}
then the first term in the expansion~\eqref{prob expand} reduces to the result~\eqref{probabilities2} for the exact probability.
\pagebreak

\pagebreak


\section{Narrow initial pointer state.  Quantitative example}
\label{Section VI} 	
In this section we will assume a Gaussian initial state for the pointer with a narrow half width~$\ell$ compared to the length of the intervals~$\Delta_{\alpha,\alpha'}$ and centered at the origin interacting with a  particle described by an initial state which is Gaussian of any half width~$\sigma$ centered at the arbitrary point~$x_0$. 

Thus at~$t=0$ the wavefunction of the particle is
\begin{equation}\label{varphi_0}
\varphi_0(x)=A{\rm e}^{-(x-x_0)^2/\sigma^2}\,,
\end{equation}
 and the normalized Gaussian describing the pointer is
 \begin{equation}\label{Phi_0}
\Phi_0(X)=\left(\frac{2}{\pi\ell^2}\right)^{1/4}{\rm e}^{-X^2/\ell^2},
\end{equation}
with~$\ell<<\delta$. 

 In the next two subsections we will examine the approximate decoherence for coarse-grained histories, as well as the approximate probabilities that can be assigned to these histories.
\subsection{Decoherence}\label{A decoherence}
 Next we insert~\eqref{varphi_0} and~\eqref{Phi_0} into{~\eqref{decoherence-functional3} and we use~\eqref{gaussian-overlap} for the integration~\eqref{pointer-overlap} over the pointer variable, to obtain for~$\alpha'=\alpha+1$
 
\begin{align}\label{|D_alpha,alpha+1|}
|D(c_\alpha,c_{\alpha+1})|\le&|A|^2|K_0|^2\mathlarger{\int}_{-\infty}^\infty {\rm d}x\nonumber\\
&\times
\mathlarger{\int}_{2\bar x_\alpha-x+\delta}^{2\bar x_{\alpha}-x+3\delta}{\rm d}x''{\rm e}^{-(x''-x_0)^2/\sigma^2}\nonumber\\
&\times
\mathlarger{\int}_{2\bar x_{\alpha}-x-\delta}^{2\bar x_{\alpha}-x+\delta}{\rm d}x'{\rm e}^{-(x'-x_0)^2/\sigma^2}\nonumber\\
&\times
{\rm e}^{-g^2(x''-x')^2/8\ell^2}.
\end{align}
 
 Then we define the new variables
 \begin{align}
 &y=\kappa x\nonumber\\
 &z=y-y_0\nonumber\\
 &w=2\bar y_\alpha-z-y_0\,,
 \end{align}
 where~$\kappa=g/\ell\sqrt{8}$, $\bar y_\alpha=\kappa\bar x_\alpha$
 and~$y_0=\kappa x_0$.
 Thus~\eqref{|D_alpha,alpha+1|} is rewritten
 \begin{align}\label{|D_alpha,alpha+1|2}
|D(c_\alpha,c_{\alpha+1})|\le&|A|^2|K_0|^2\frac{1}{\kappa^3}\mathlarger{\int}_{-\infty}^\infty {\rm d}w\nonumber\\
&\times
\mathlarger{\int}_{w+\gamma}^{w+3\gamma}{\rm d}z''\mathlarger{\int}_{w-\gamma}^{w+\gamma}{\rm d}z'\nonumber\\
&\times
{\rm e}^{-z''^2/\bar\alpha^2\sigma^2}{\rm e}^{-z'^2/\kappa^2\sigma^2}\nonumber\\
&\times
{\rm e}^{-(z''-z')^2},
\end{align}
 and
 \begin{equation}\label{gamma}
 \gamma=\kappa\delta\,.	
 \end{equation}
In the overlap integral~\eqref{gaussian-overlap} the main contribution arises from the region
\begin{equation}
|x''-x'|\le \frac{\sqrt{8}\ell}{g}, 
 \end{equation}
 where for a given~$x$ in~\eqref{|D_alpha,alpha+1|},~$(x'+x)/2\in \Delta_\alpha$ and~$(x''+x)/2\in\Delta_{\alpha+1}$.  Since~$\ell<<\delta$ it follows that~$(x'+x)/2$ is close to the upper boundary of~$\Delta_\alpha$ and~$(x''+x)/2$ is very close to the lower boundary of~$\Delta_{\alpha+1}$ as shown in Fig~\ref{fig2}.  In terms of the variables~$w$,~$z'$ and~$z''$ this is shown in Fig \ref{fig6}.  That is,~$z'$ and~$z''$ are very close to the value~$w+\gamma$ which corresponds to the common boundary of the two contiguous regions~$\Delta_\alpha$ and~$\Delta_{\alpha+1}$.  Thus~$z'\lesssim w+\gamma$ and~$z''\gtrsim w+\gamma$.
 Next we expand~${\rm e}^{-z''^2/\kappa^2\sigma^2}{\rm e}^{-z'^2/\kappa^2\sigma^2}$ about~$w+\gamma$, and when we keep the first two terms in the expansion in~\eqref{|D_alpha,alpha+1|2} we obtain
\begin{align}\label{|D_alpha,alpha+1|3}
|D(c_\alpha,c_{\alpha+1})|&\le|A|^2|K_0(\omega,T)|^2\frac{1}{\kappa^3}\mathlarger{\int}_{-\infty}^\infty {\rm d}w\nonumber\\
&\times
\mathlarger{\int}_{w+\gamma}^{w+3\gamma}\!\!\!\!\!\!{\rm d}z''\mathlarger{\int}_{w-\gamma}^{w+\gamma}\!\!\!\!{\rm d}z'{\rm e}^{-(z''-z')^2}{\rm e}^{-2(w+\gamma)^2/\kappa^2\sigma^2}\nonumber\\
&\times
\left\{1-\frac{1}{\kappa^2\sigma^2}\left[z''^2+z'^2-2(w+\gamma)^2 \right]+\dots \right\}
\end{align}
\pagebreak\\
or
\begin{equation}\label{|D|-expansion2}
|D(c_\alpha,c_{\alpha+1)}|\le D^{(0)}c_\alpha,c_{\alpha+1})+D^{(1)}(c_\alpha,c_{\alpha+1}),
\end{equation}
with
\begin{align}\label{D(0)}
D^{(0)}(c_\alpha,c_{\alpha+1}&)=A|^2|K_0(\omega,T)|^2\frac{1}{\kappa^3}\mathlarger{\int}_{-\infty}^\infty {\rm d}w\nonumber\\
&\times
\mathlarger{\int}_{w+\gamma}^{w+3\gamma}{\rm d}z''\mathlarger{\int}_{w-\gamma}^{w+\gamma}{\rm d}z'{\rm e}^{-(z''-z')^2},
\end{align}
and
\begin{align}\label{D(2)}
D^{(1)}(c_\alpha,c_{\alpha+1})&=A|^2|K_0|^2\frac{1}{\kappa^5\sigma^2}\mathlarger{\int}_{-\infty}^\infty {\rm d}w\,{\rm e}^{-2(w+\gamma)^2/\kappa^2\sigma^2}\nonumber\\
&\times
\mathlarger{\int}_{w+\gamma}^{w+3\gamma}{\rm d}z''\mathlarger{\int}_{w-\gamma}^{w+\gamma}{\rm d}z'{\rm e}^{-(z''-z')^2}\nonumber\\
&\times
\left[-z''^2-z'^2+2(w+\gamma)^2 \right].
\end{align}
In the lowest order term~\eqref{D(0)} let
\begin{equation}\label{F}
F_{\alpha,\alpha+1}= \mathlarger{\int}_{w+\gamma}^{w+3\gamma}{\rm d}z''\mathlarger{\int}_{w-\gamma}^{w+\gamma}{\rm d}z'{\rm e}^{-(z''-z')^2}.
\end{equation}
Then the evaluation of~\eqref{F} yields the~$w$-independent result
\begin{align}
F_{\alpha,\alpha+1}=& \frac{1}{2}\Big\{ 1+{\rm e}^{-16\gamma^2}
-2{\rm e}^{-4\gamma^2}\nonumber\\
&
+4\sqrt{\pi}\gamma\left[  -{\rm erf}(2\gamma)+{\rm erf}(4\gamma)\right]\Big\}.
\end{align}
Thus after evaluating the remaining integration over~$w$ we obtain
\begin{equation}
D^{(0)}(c_\alpha,c_{\alpha+1})=\sqrt{\frac{\pi}{2}}|K_0|^2|A|^2\frac{\sigma}{\kappa^2} F_{\alpha,\alpha+1},
\end{equation}
where
\begin{align}
F_{\alpha,\alpha+1}=&\frac{1}{2}\Big\{ 1+{\rm e}^{-16\gamma^2}-2{\rm e}^{-4\gamma^2}\nonumber\\
&
+4\sqrt{\pi}\gamma\left[  -{\rm erf}(2\gamma)+{\rm erf}(4\gamma)\right]\Big\}.
\end{align}
Similarly in the higher order term~$D^{(1)}$ in~\eqref{D(2)} let
\begin{align}\label{G}
G_{\alpha,\alpha+1}=&\mathlarger{\int}_{w+\gamma}^{w+3\gamma}{\rm d}z''\mathlarger{\int}_{w-\gamma}^{w+\gamma}{\rm d}z'{\rm e}^{-(z''-z')^2}\nonumber\\
&\times
 \left[-z''^2-z'^2+2(w+\gamma)^2\right].
 \end{align}
 Then the integrations over the variables~$z'$ and~$z''$ yield once more a~$w$-independent result
\begin{align}
G_{\alpha,\alpha+1}&=
 \frac{1}{3}\Big\{ 1+(1+4\gamma^2){\rm e}^{-16\gamma^2}
 -2(1+4\gamma^2){\rm e}^{-4\gamma^2}\nonumber\\
&+\sqrt{\pi}\gamma(3+16\gamma^2)\left[  -{\rm erf}(2\gamma)+{\rm erf}(4\gamma)\right]\Big\}.
\end{align}
 Thus the higher order term~\eqref{D(2)} can be evaluated to yield
 \begin{equation}
D^{(1)}(c_\alpha,c_{\alpha+1})=\sqrt{\frac{\pi}{2}}|K_0|^2|A|^2\frac{1}{\kappa^4\sigma} G_{\alpha,\alpha+1}\,.
\end{equation}
For~$\gamma\geqslant\frac{1}{2}$ we have the asymptotic limits~$F_{\alpha,\alpha+1}\rightarrow 1/2$ and~$G_{\alpha,\alpha+1}\rightarrow 1/3$. Inserting these values into~\eqref{|D|-expansion2} we finally obtain the decoherence condition
\begin{equation}
|D(c_\alpha,c_{\alpha+1)}|\le |K_0|^2\frac{4\ell^2}{g^2}\left(1-\frac{16}{3g^2}\frac{\ell^2}{\sigma^2}+\dots\right)
\end{equation}
where~$16\ell^2/3g^2\sigma^2<1$, in addition to~$\ell<<\delta$.
With
\begin{equation}
|K_0|^2=\frac{m\omega}{2\pi \sin\omega T},
\end{equation}
the lowest order term for~$D(c{_\alpha},c_{\alpha+1})$ is
\begin{equation}
D^{(0)}(c_\alpha,c_{\alpha+1})=\frac{2m\omega \ell^2}{\pi g^2(\omega,T)\sin\omega T}.
\end{equation}
This exact expression for~$D^{(0)}(c_\alpha,c_{\alpha+1})$ has an extra factor of two compared to the semi- qualitative result obtained in~\eqref{pointer-decoherene}.


	\begin{figure}[h] 
	\psset{unit=.85cm}
	\begin{pspicture}(-2,-1.5)(2,1.5)
	%
	\psline[linewidth=1pt](-3.5,0)(3.5,0)
	\psline[linestyle=dashed,linewidth=.5pt,linecolor=blue] (0,-1)(0,1)
	\psline[linewidth=0.5pt] (1.75,-.25)(1.75,.25)
	\psline[linewidth=0.5pt] (-1.75,-.25)(-1.75,.25)
	\psline[linewidth=0.5pt] (3.5,-.25)(3.5,.25)
	\psline[linewidth=0.5pt] (-3.5,-.25)(-3.5,.25)

	\rput[bh](-1.75,0.4){$\mathsmaller{w}$}
	\rput[bh](1.75,0.4){$\mathsmaller{w+2\gamma}$}
	\rput[bv](-0.4,0.2){$\mathsmaller{z'}$}
	\rput[bv](0.5,0.2){$\mathsmaller{z''}$}

	\psline[linewidth=0.5pt] (.5,-.10)(.5,.10)
	\psline[linewidth=0.5pt] (-.4,-.10)(-.4,.10)

	\rput[bh](3.5,0.4){$\mathsmaller{w+3\gamma}$}
	\rput[bh](-3.5,0.4){$\mathsmaller{w-\gamma}$}

	\rput[bc](0.1,-1.2){$\mathsmaller{w+\gamma}$}
  
	\end{pspicture}
	\caption{The relative position of~$z'$ and~$z''$
  	in the contiguous intervals corresponding to~$\Delta_\alpha$ and~$\Delta_{\alpha+1}$. The vertical 		dashed line is the common boundary. The length of each interval is~$2\gamma$, with~$\gamma=\frac	{g}{\sqrt{8}}\frac{\delta}{\ell}$. The values~$z'$ and~$z''$ are very close to each other and on 		either side of the 	boundary at~$w+\gamma$.}
	\label{fig6}
	\end{figure}
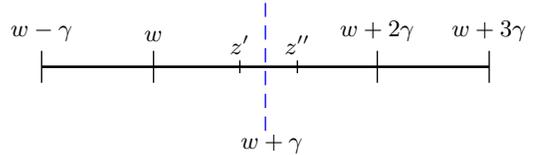
\pagebreak
\subsection{Probabilities}\label{B probabilities}
Next we show the results for approximate probabilities.  We will consider two cases for the normalization of the initial wavefunction for the pointer
\begin{equation*}
\Phi_0(X)=B{\rm e}^{-X^2/\ell^2}\,.
\end{equation*}

a)  Normalized pointer:
\begin{equation*}
\int_{-\infty}^\infty|\Phi_0(X)|^2{\rm d}X=1
\end{equation*}
\begin{equation}
B=\left(\frac{2}{\pi \ell^2}\right)^{1/2}
\end{equation}

b)  The pointer is normalized by the condition
\begin{equation*}
\lim_{\ell \to 0}\Phi_0(X)= \delta(X)
\end{equation*}

\begin{equation}
B=\frac{1}{\sqrt{\pi} \ell}
\end{equation}
We expand in a similar fashion as was done in~\eqref{|D_alpha,alpha+1|2}.
Keeping the lowest order term in the expansion we obtain
\begin{align}\label{prob-functional}
&|D(c_\alpha,c_\alpha)|\leq|K_0|^2|A|^2\frac{1}{\kappa^3}\int_{\infty}^\infty {\rm d}w\int_{w-\gamma}^{w+\gamma}{\rm d}z"\nonumber\\
&\times
\int_{w-\gamma}^{w+\gamma}{\rm d}z'{\rm e}^{-(z"^2+z'^2)/\lambda^2}
|B|^2{\rm e}^{-(z"-z')^2}.
\end{align}
with~$\lambda=\kappa\sigma$, and~$\kappa$ is defined in~\eqref{kappa}.

Next, since the pointer is in a narrow initial state, we have~$z'\approx z''$. We can expand~${\rm e}^{-(z''^2+z'^2)/\lambda^2}$ about~$z''=z'$ and we obtain for the lowest order term
\begin{equation}
|D(c_\alpha,c_\alpha)|\leq|K_0|^2|A|^2|B|^2\frac{1}{\kappa^3}\int_{\infty}^\infty{\rm d}wI(w),
\end{equation}
where
\begin{equation}
I(w)=\int_{w-\gamma}^{w+\gamma}{\rm d}z"
\int_{w-\gamma}^{w+\gamma}{\rm d}z'{\rm e}^{-2z''^2/\lambda^2}
{\rm e}^{-(z"-z')^2}\,.
\end{equation}
The integration over~$z,$ can easily be performed to obtain
\begin{align}
I(w)&=\frac{1}{2}\sqrt{\pi}\int_{w-\gamma}^{w+\gamma}{\rm d}z''
\int_{w-\gamma}^{w+\gamma}{\rm d}z'{\rm e}^{-2z''^2/\lambda^2}\nonumber\\
&\times
\big[{-\rm erf}(w-z''-\gamma)
+{\rm erf}(w-z''+\gamma)\big]\,.
\end{align} 
Since the pointer is narrow we have~$\ell\ll\delta$ and thus in~\eqref{gamma}~$\gamma\gg 1$;
therefore we can approximate
\begin{align}
{-\rm erf}&(w-z''-\gamma)
+{\rm erf}(w-z''+\gamma)\approx\nonumber\\
&
2\big[-\theta(w-z''-\gamma)+\theta(w-z''+\gamma)\big],
\end{align}
where~$\theta(w-z''-\gamma)$ and~$\theta(w-z''+\gamma)$ are two unit step functions.
\pagebreak
This produces the following result for the integration~$I(w)$:
\begin{equation}
I(w)= 2\sqrt{2\pi}\gamma\lambda.
\end{equation}

 For the lowest order term we obtain for the probabilities~$p_\alpha$
\begin{equation}
p_\alpha\leq
\begin{cases}
~~\frac{4}{g(\omega,T)}|K_0|^2\sqrt{2\pi}\ell\delta\quad{\rm for~case~(a)} \\\\
~~\frac{4}{g(\omega,T)}|K_0|^2\delta\quad{\rm for~case~(b)}\,. 
\end{cases}
\end{equation}
The result for (b) agrees with~\eqref{probabilities4} as expected, and in this case the equality holds.

Higher order terms can be evaluated by expanding~${\rm e}^{-(z'^2+z''^2)/\lambda^2}$ in powers of~$z''-z'$:
\begin{align}
{\rm e}^{-(z'^2+z''^2)/\lambda^2}=&\,{\rm e}^{-2z'^2/\lambda^2}-\frac{2}{\lambda^2}{\rm e}^{-2z'^2/\lambda^2}z'(z''-z')\nonumber\\
+&
\frac{1}{\lambda^4}{\rm e}^{-2z'^2/\lambda^2}(2z'^2-\lambda^2)(z''-z')^2\nonumber\\
+&
O\left((z''-z')^3\right)\,.
\end{align}
Inserting into~\eqref{prob-functional} we obtain that the~$O(z''-z')$ term vanishes when~$\gamma$ is large. The next non-vanishing term is~$O\left((z'-z'')^2\right)$. 

\begin{figure}
	\centerline{\epsfxsize=3.5in\epsfbox{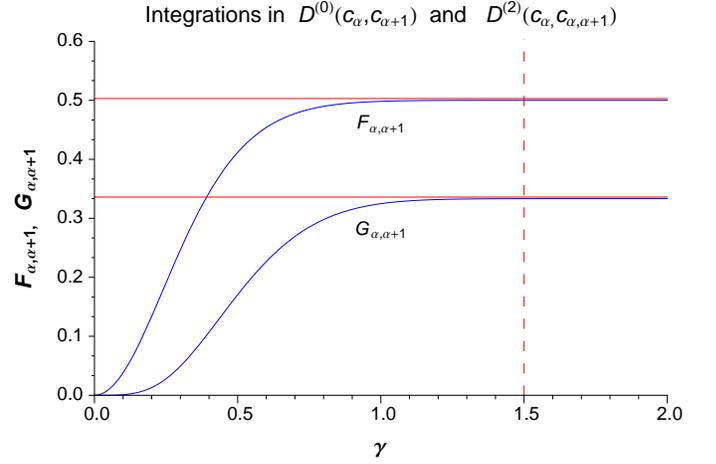}}
	\caption{The result of the integrations in~\eqref{F} and~\eqref{G} for the case of a narrow pointer.  The solid red lines are 	the asymptotes at~$F_{\alpha,\alpha+1}=1/2$ and~$G_{\alpha,\alpha+1}=1/3$.  The 	vertical red dotted line is at~$ \gamma=1.5$ for which~$F_{\alpha,\alpha+1}=0.50$ and~$G_{\alpha,	\alpha+1}=0.33$.}
	 \label{fig7}
	\end{figure}
\pagebreak
	
\section{Summary and Conclusion}
\label{conclusion}	
We have investigated the decoherence of coarse-grained histories of a particle coupled to a von Neumann apparatus.   As posited in Ref.~\cite{H2}, realistic measurement situations must take into account the finite time over which they take place, as well as the localization of the particle to a position interval~$\Delta$.  This naturally has lead to the consideration of spacetime coarse grainings in the present paper.

The particle under consideration is a harmonic oscillator acted on by a driving force. In the appropriate limits, we can describe both a free particle as well as a particle acted on by a force as particular cases of the formalism for the driven oscillator. We have studied the system particle-apparatus in the context of the Gell-Mann and Hartle quantum mechanics of closed systems. The coarse-grained histories that we have considered are such that
the real line of all values for the average position~$\bar x$ on a given Feynman path for the particle has been divided into equal length intervals:
\begin{equation*}
\Delta_{\alpha}=\big(\,\bar x_{\alpha}-\delta/2\,,\,\bar x_{\alpha}+\delta/2\,\big],~~{\rm with}~\delta>0.
\end{equation*}
A class of particle paths~$c_{\alpha}$ has been defined by the set of all paths~$x(t)$ on the time interval~$[0,T]$  such that on any given path we get the arithmetic average of the initial and final position of the particle~$\bar x\in \Delta_{\alpha}$.
Therefore  the set~$C$ of all paths in the time interval~$[0,T]$  has been divided into an exhaustive set of mutually exclusive classes~$c_{\alpha}$. For each class the von Neumann apparatus correlates with the arithmetic average~$\bar x\in \Delta_\alpha$ of the initial and final position of each path. 

 Decoherence depends on the initial state of the system particle-apparatus.  If the particle is decoupled from the apparatus then there is exact decoherence when the particle starts in a sharp value of the position.  When the particle is coupled to the apparatus the same result follows.  Meanwhile the apparatus has been assumed to be in any normalizable initial state.  Relative probabilities can be assigned in this case, and it is found that the probability for each coarse-grained history is the same, regardless of the class, and that they are proportional to the length of the intervals~$\Delta_\alpha$ for each class.  In addition the driving force acting on the particle oscillator does not affect the result for the probabilities. These probabilities are relative and do not add up to unity but instead add up to infinity because the initial state of the system is not normalizable.  When the particle is not coupled to the apparatus the same probabilities are obtained. 
 
 There is also exact decoherence when the initial state is a sharp state of position for the pointer and the particle is an any normalized state.  In this case relative probabilities can be assigned to the coarse-grained histories.  These probabilities are also proportional to the length of the intervals~$\Delta_\alpha$ and inversely proportional to the coupling constant ~$g(\omega,T)$ that appears in the shift in the position of the pointer at the end of the measurement. 
 
Next, approximate decoherence is considered for the case of a narrow initial particle state.   Semi quantitative arguments reveal that there will be approximate decoherence if the initial state of the particle is narrow compared with the width~$\delta$ of the intervals~$\Delta_\alpha$.  For a normalized initial particle state of narrow width the absolute value of the decoherence functional is proportional to the square of the small width~$\sigma$ of such state.  In the limiting case that the state reduces to an eigenstate of position when the width of the state vanishes, we obtain that the absolute value of the decoherence functional is proportional to the width~$\sigma$ of the initial state.  When a detailed calculation is done  starting with a pure initial  state consisting of  a narrow Gaussian state for the particle and a Gaussian state for the pointer such that~$\sigma\ll\frac{\sqrt{2}}{g}\ell$, where~$\ell$ is the width of the pointer, we have obtained the first two terms in an expansion of powers of~$\frac{\sigma}{\ell}$.  Approximate probabilities can also be assigned and have been obtained as a power expansion  similar to the one applied to the decoherence functional.  In the limit~$\ell\rightarrow 0$ we recover the probabilities obtained earlier for the case of sharp decoherence.  These probabilities have been evaluated for the semi-qualitative calculation as well as for the quantitative example.

In the same vein approximate decoherence is satisfied when the initial state of the pointer has a half width~$\ell$ which is much smaller compared to the width~$\delta$ of the intervals~$\Delta_\alpha$. Semi-qualitative arguments are provided for the case of a narrow Gaussian pointer state and any normalized initial particle state. Likewise we have developed a detailed calculation with a narrow Gaussian initial state of the pointer and an initial Gaussian state of the particle centered at some arbitrary point~$x_0$ and arbitrary width. We obtain that approximate decoherence is satisfied if the width of the initial state of the particle is larger than the width of the narrow Gaussian pointer according to the inequality~$\sigma\gg\frac{4}{\sqrt{3}g}\ell$ in addition to~$\ell\ll\delta$.  Approximate probabilities can be assigned and a power series expansion produces a lowest order term that coincides with the probabilities for the case of exact decoherence in the limit~$\ell\rightarrow 0$ when the initial state of the pointer is an eigenstate of position at~$X=0$. 


To conclude, apart from having studied decoherence for coarse-histories of a driven oscillator coupled to a von Neumann apparatus, the current work has exhibited some of the difficulties entailed in calculating the decoherence functional save for the most simple situations.  


\begin{acknowledgments}
I am grateful to James B. Hartle for useful conversations during the early stages of this paper and to Michael P. McLaughlin for useful consultations regarding the numerical work.
\end{acknowledgments}

	
\appendix\label{AppendixA}

\renewcommand{\theequation}{\Alph{section}.\arabic{equation}}
\section{Property of the decoherence functional}
\label{appendixA}
In this appendix we proceed to prove that Eq.\eqref{decoherence-functional3} for the decoherence functional satisfies the general condition
\begin{equation}\label{property2}
\sum_{\alpha,\alpha'}D(c_\alpha,c_{\alpha'})=\Braket{\Psi_0|\Psi_0}.
\end{equation}
Inserting~(\ref{decoherence-functional3}) into~(\ref{property2}) we obtain
\begin{align}\label{Sum-decoherence-functional}
\sum_{\alpha,\alpha'}D(c_\alpha,c_{\alpha'})&=\mathlarger{\int}_{-\infty}^\infty \sum_{\alpha'}
\mathlarger{\int}_{2\bar x_{\alpha'}-x-\delta}^{2\bar x_{\alpha'}-x+\delta}\!\!\!\!{\rm d}x''\sum_\alpha
\mathlarger{\int}_{2\bar x_{\alpha}-x-\delta}^{2\bar x_{\alpha}-x+\delta}\!\!\!\!{\rm d}x'\nonumber\\
&\times
K_0^{*}(x,T;x'',0)K_0(x,T;x',0)\nonumber\\
&\times
\mathlarger{{\rm e}}^{\mathlarger{\mathlarger{-\rm i}\phi\left(\frac{x+x''}{2},\omega,T\right)}}\mathlarger{{\rm e}}^{\mathlarger{\mathlarger{\rm i}\phi\left(\frac{x+x'}{2},\omega,T\right)}}\nonumber\\
&\times
\varphi^{*}_0(x'')\varphi_0(x')\mathlarger{\mathlarger{\int}}_{-\infty}^\infty{\rm d}X'\Phi_0(X')\nonumber\\
&\times
\Phi_0^{*}\left(X'-g(\omega,T)\frac{x''-x'}{2}\right).
\end{align}
Next from the properties 
\begin{align}
\Delta_{\alpha}\bigcap \Delta_{\alpha'}&=\O,~~{\rm with}~\alpha\not =\alpha'\nonumber\\
\bigcup_{\alpha\in \mathbb{Z}}\Delta_{\alpha}&=(-\infty,\infty)
\end{align}
and
\begin{equation}
K_0(x,T;x',0)=\Braket{x|\hat U_{\mathit{particle}}|x'},
\end{equation}
where
\begin{equation}
\hat U_{\mathit{particle}}=\exp\left[-{\rm i}\left(\frac{p^2}{2m}+\frac{m}{2}\omega^2x^2\right)T\right],
\end{equation}
the decoherence functional~\eqref{decoherence-functional4} is rewritten
\begin{align}\label{sum-decoherence-functional5}
\sum_{\alpha,\alpha'}D(c_\alpha,c_{\alpha'})=&
\mathlarger{\mathlarger{\iiint}}\limits_{\!\!\!\!-\infty}^{\,,\,\,\,\,\,\,\infty}{\rm d}x\,{\rm d}x'\,{\rm d}x''\Braket{x''|\hat U^{\dagger}_{\mathit{particle}}|x}\nonumber\\
&\times
\Braket{x|\hat U_{\mathit{particle}}|x'}\mathlarger{{\rm e}}^{\displaystyle{-\rm i}\bar\phi(x''-x')}\nonumber\\
&\times
\varphi^{*}_0(x'')\varphi_0(x')\Delta_{\mathit{overlap}}(x''-x'),
\end{align}
with
\begin{align}
\bar\phi(x''-x')=&-\phi\left(\frac{x+x''}{2},\omega,T\right)\nonumber\\
&
+\phi\left(\frac{x+x'}{2},\omega,T\right)\,.
\end{align}
That is,~$\bar\phi$ does not depend on~$x$ and only on the difference~$x''-x'$.  This follows readily from the structure of the phase~$\phi$ defined by  Eq.(34) in~\cite{FR-1}.  The factor~$\Delta_{\mathit{overlap}}(x''-x')$ is defined in~(\ref{pointer-overlap}) which also depends on the difference~$x''-x'$.
From here it is easily seen that~\eqref{sum-decoherence-functional5} collapses to~$\Braket{\Psi_0|\Psi_0}$.


\section{Decoherence and approximate probabilities for Gaussian initial particle states}\label{AppendixB}

We start with the lowest order term~$D^{(0)}(c_\alpha,c_{\alpha'})$ in~\eqref{|D|-expansion}.  In expression~\eqref{|D|^(0)} we change to a dimensionless form for the integrations.  That is, introduce the dimensionless variables~$y=\frac{x}{\sigma}$,~~$\bar y_\alpha=\frac{\bar x_\alpha}{\sigma}$, and~~$y_0=\frac{x_0}{\sigma}$.  We further let~$z=y-y_0$ and~$w=2\bar y_\alpha-z-y_0$ to obtain
\begin{align}
D^{(0)}(c_\alpha,c_{\alpha'})=&|K_0|^2|A|^2\sigma^3\int_{-\infty}^\infty{\rm d}w\int_{w-\beta}^{w+\beta}{\rm d}z'{\rm e}^{-z'^2}\nonumber\\
&\times
\int_{2\Delta_{\alpha\alpha'}+w-\beta}^{2\Delta_{\alpha\alpha'}+w+\beta}{\rm d}z''{\rm e}^{-z''^2},
\end{align}
with~~$\beta=\frac{\delta}{\sigma}$~~~and~~~$\Delta_{\alpha\alpha'}=n\beta,~~n=1,2,3\dots$
The integrations over~$z'$ and~$z''$ are easily carried out and
\begin{align}\label{B2}
D^{(0)}(c_\alpha,c_{\alpha'})=&|K_0|^2|A|^2\sigma^3\int_{-\infty}^\infty{\rm d}w\nonumber\\
&\times
\frac{\pi}{4}\left[{\rm erf}(w+\beta)-{\erf}(w-\beta)\right]\nonumber\\
&\times
\Big[{\rm erf}(w+2\Delta_{\alpha\alpha'}+\beta)\nonumber\\
&
-{\rm erf}(w+2\Delta_{\alpha\alpha'}-\beta)\Big]\,.
\end{align}
The integration over~$w$ can be done using the general result for the product of two error functions
\begin{align}\label{B3}
&\int_{-\infty}^\infty{\rm d}x\Big[{\rm erf}(a+x)\,{\erf}(b-x)+1\Big]=\nonumber\\
&
2(a+b)\,{\erf}\left(\frac{a+b}{\sqrt 2}\right)+2\sqrt{\frac{2}{\pi}}{\rm e}^{-(a+b)^2/2}
\end{align}
plus
\begin{equation}
\int_{-\infty}^\infty{\rm d}x\Big[{\erf\,^2(x+\beta})-1\Big]=-2\sqrt\frac{2}{\pi}\,.
\end{equation}
Then the integration in~\eqref{B2} when~$\alpha'=\alpha+1$ yields
\begin{align}\label{evalB2} 
\frac{\pi}{4}&\int_{-\infty}^\infty {\rm d}w\left[{\erf}(w+\beta)-{\erf}(w-\beta)\right]\nonumber\\
&\times
\left[{\erf}(w+3\beta)-{\erf}(w+\beta)\right]\nonumber\\
&=
-2\pi\beta\,{\erf}(\sqrt{2}\beta)+2\pi\beta\,{\erf}(2\sqrt{2}\beta)\nonumber\\
&
-\sqrt{2\pi}{\rm e}^{-2\beta^2}+\frac{1}{2}\sqrt{2\pi}{\rm e}^{-8\beta^2}+\frac{1}{2}\sqrt{2\pi}
\end{align}
which is expression~\eqref{I_a,a+1} for~$I_{\alpha,\alpha+1}$.
\pagebreak
The evaluation of~$D^{(1)}(c_\alpha,c_{\alpha'})$ follows in a similar fashion:
\begin{align}\label{B6}
D^{(1)}(c_\alpha,c_{\alpha'})=&-\kappa^2|K_0|^2|A|^2\sigma^5\int_{-\infty}^\infty{\rm d}w\nonumber\\
&\times
\int_{2\Delta_{\alpha\alpha'}+w-\beta}^{2\Delta_{\alpha\alpha'}+w+\beta}{\rm d}z''{\rm e}^{-z''^2}\nonumber\\
&\times
\int_{w-\beta}^{w+\beta}{\rm d}z'(z''-z')^2{\rm e}^{-z'^2}\,.
\end{align}
The integrations over~$z'$ and~$z''$ in~\eqref{B6} yield, after shifting the~$w$ variable according to~$w\rightarrow w-\beta,$ and with~$\alpha'=\alpha+1$
\par\vspace{-3ex}
\begin{align}
&
\int_w^{w+2\beta}{\rm d}z''
\int_{w-2\beta}^w{\rm d}z'(z''-z')^2{\rm e}^{-(z'^+z''^2)}\nonumber\\
&=A+B,
\end{align}
where~$A$ and~$B$ are given by the expressions
\begin{widetext}
\begin{align}
A=&\frac{1}{2} e^{-2 w^2}+\frac{1}{2} e^{-2 w^2-8 \beta ^2}-\frac{1}{2} e^{-2 w^2-4 w \beta -4 \beta ^2}-\frac{1}{2} e^{-2 w^2+4 w
\beta -4 \beta ^2}-\frac{1}{4} e^{-w^2-4 w \beta -4 \beta ^2} \sqrt{\pi } w\, \text{erf}(w)-\nonumber\\
&
\frac{1}{4} e^{-w^2+4 w \beta -4 \beta ^2} \sqrt{\pi } w \,\text{erf}(w)-\frac{1}{2} e^{-w^2-4 w \beta -4 \beta ^2} \sqrt{\pi } \beta \, \text{erf}(w)+\frac{1}{2}
e^{-w^2+4 w \beta -4 \beta ^2} \sqrt{\pi } \beta \, \text{erf}(w)-\nonumber\\
&
\frac{1}{4} \pi \, \text{erf}(w)^2+\frac{1}{4} e^{-w^2-4 w \beta -4 \beta ^2} \sqrt{\pi } w \,\text{erf}(w-2 \beta )+\frac{1}{2} e^{-w^2-4 w \beta
-4 \beta ^2} \sqrt{\pi } \beta \, \text{erf}(w-2 \beta )+\nonumber\\
&
 \frac{1}{4} e^{-w^2+4 w \beta -4 \beta ^2} \sqrt{\pi } w\, \text{erf}(w+2 \beta )-\frac{1}{2} e^{-w^2+4 w \beta -4 \beta ^2} \sqrt{\pi } \beta
\, \text{erf}(w+2 \beta )-\nonumber\\
&
\frac{1}{2} e^{-w^2} \sqrt{\pi } w \,\text{erfc}(w)+\frac{1}{4} e^{-w^2} \sqrt{\pi } w \,\text{erfc}(w-2 \beta )+\frac{1}{4} e^{-w^2} \sqrt{\pi
} w\, \text{erfc}(w+2 \beta )
\end{align}
\end{widetext}
and
\begin{align}
B=&\frac{\pi}{4}\Big[\text{erf}(w)\,\text{erf}(w-2\beta)
+
\text{erf}(w)\,\text{erf}(w+2\beta)\nonumber\\
&
-\text{erf}(w-2\beta)\,\text{erf}(w+2\beta)\Big]\,.
\end{align}
With the aid of~\eqref{B3} we obtain
\begin{align}
\int_{-\infty}^\infty\left(B-\frac{\pi}{4}\right) \, {\rm d}w=& 2 e^{-8 \beta ^2} \sqrt{\frac{2}{\pi }}-4 e^{-2 \beta ^2} \sqrt{\frac{2}{\pi }}\nonumber\\
&
-8 \beta  \,\text{erf}\left(\sqrt{2}
\beta \right)+8 \beta \, \text{erf}\left(2 \sqrt{2} \beta \right)\,.
\end{align} 
The term~$\frac{\pi}{4}$ has been subtracted from~$B$ in order to facilitate the integration using the result~\eqref{B3}.  A similar term is added to~$A$ and we evaluate the integral
\begin{align}
\int_{-\infty}^\infty \left(A+\frac{\pi}{4}\right){\rm d}w=&e^{-8 \beta ^2} \sqrt{\frac{\pi }{2}}+\sqrt{2 \pi }\nonumber\\
&
-e^{-2 \beta ^2} \sqrt{2 \pi}. 
\end{align}
Thus the integrations in~\eqref{B6} yields the result
\begin{align}\label{evalB12}
\int _{-\infty }^{\infty }{\rm d}w&\int _{w+\beta }^{w+3 \beta }{\rm d}z''\int _{w-\beta }^{w+\beta }{\rm d}z'\nonumber\\
&\times
(z''-z')^2e^{-\left(z'^2+z''^2\right)}=\nonumber\\
&
\left(1+e^{-8 \beta ^2}-2 e^{-2 \beta ^2}\right) \sqrt{2 \pi }\nonumber\\
&
+2 \pi  \beta  \left[-\text{erf}\left(\sqrt{2} \beta \right)+\text{erf}\left(2
\sqrt{2} \beta \right)\right],
\end{align}
which is expression~\eqref{J_a,a+1} for~$J_{a,a+1}$.

The expression~\eqref{approxprob}for the approximate probabilities is rewritten in terms of dimensionless integrations
\begin{align}\label{B13}
p_{\alpha}\lesssim&\,|K_0|^2|A|^2|\sigma^3\int_{-\infty}^\infty {\rm d}w\int_{w-\beta}^{w+\beta}{\rm d}z''\int_{w-\beta}^{w+\beta}{\rm d}z'\nonumber\\
&\times
{\rm e}^{-(z''^2+z'^2)}\left[1-\kappa^2(z''-z')^2+\dots\right]\,.
\end{align}
The lowest order term in the expansion in powers of~$\alpha$ yields the integral
\begin{align}\label{B14}
\int_{-\infty}^\infty{\rm d}w\left(\int_{w-\beta}^{w+\beta}{\rm d}z\,{\rm e}^{-z^2}\right)^2=&
-\sqrt{2\pi}+\sqrt{2\pi}{\rm e}^{-2\beta^2}\nonumber\\
&
+2\pi\beta\,\text{erf}(\sqrt{2}\beta)\,.
\end{align}

The next higher order term can be evaluated the same way as in~\eqref{evalB2} and~\eqref{evalB12}.  This yields the integral 
\begin{align}\label{B15}
\int_{-\infty}^\infty &{\rm d}w\int_{w-\beta}^{w+\beta}{\rm d}z''\int_{w-\beta}^{w+\beta}{\rm d}z'\nonumber\\
&\times
(z''-z')^2{\rm e}^{-(z''^2+z'^2)}=\nonumber\\
&\sqrt{2\pi}(-1+{\rm e}^{-2\beta^2})+2\pi\beta\text{erf}(\sqrt{2}\beta).
\end{align}
Putting together~\eqref{B14} and~\eqref{B15} into~\eqref{B13} we get the result
\begin{align}
p_\alpha\lesssim&|K_0|^2|A|^2\sigma^3\bigg[\sqrt{2\pi}(-1+e^{-2\beta^2})\nonumber\\
&+2\pi\beta\text{erf}(\sqrt{2}\beta)\bigg]
\left[1-\frac{g^2}{8}\frac{\sigma^2}{\ell^2}+\dots\right].
\end{align}
In the limit of large~$\beta$ we get
\begin{equation}
p_\alpha\lesssim|K_0|^2|A|^2\sigma^3
2\pi\beta
\left[1-\frac{g^2}{8}\frac{\sigma^2}{\ell^2}+\dots\right].
\end{equation}



%
%
%
%
%
%
%
%
%
%

\bibliography{DCH_hist}

\end{document}